\documentclass[amsmath,amssymb,showpacs,showkeywords,twocolumn]{revtex4}
\usepackage[dvips]{graphicx,color}
\usepackage{times}
\usepackage{mathrsfs}
\usepackage{amsmath}
\usepackage{graphicx}
\usepackage{epsfig}
\usepackage{dcolumn}
\usepackage{bm}
\usepackage{multirow}
\begin{document}
\title{Ballistic transport and spin dependent anomalous quantum tunnelling in Rashba-Zeeman and bilayer graphene hybrid structures}
\author{Saumen Acharjee\footnote{saumenacharjee@dibru.ac.in},  Arindam Boruah\footnote{arindamboruah@dibru.ac.in}, Reeta Devi\footnote{reetadevi@dibru.ac.in} and Nimisha Dutta\footnote{nimishadutta@dibru.ac.in}}
\affiliation{Department of Physics, Dibrugarh University, Dibrugarh 786 004, 
Assam, India,}

\begin{abstract}
In this work, we have studied the spin-dependent ballistic transport and anomalous quantum tunnelling in Bilayer Graphene (BLG) hybrid connected to two Rashba-Zeeman (RZ) leads under an external electric biasing. We investigated the transmission and conductance for the proposed system using scattering matrix formalism and Landauer - B\"{u}ttiker formula considering a double delta-like barrier under a set of experimentally viable parameters. We found that the transmission characteristics are notably different for up and down spin incoming electrons depending upon the strength of magnetization. Moreover, the transmission of up and down spin electrons is found to be magnetization orientation dependent. The maximum tunnelling and conductance can be achieved by tuning biasing energy and magnetization strength and choosing a material with suitable Rashba Spin-Orbit Coupling (RSOC). This astonishing property of our system can be utilized in fabricating devices like spin filters. We found the Fano factor of our system is $0.4$ under strong magnetization conditions while it reduces to $0.3$ under low magnetization conditions. Moreover, we also noticed that the transmission and conductance significantly depend on the Rashba - Zeeman effect. So, considering a suitable RZ material, the tunnelling of the electrons can be tuned and controlled. 
\end{abstract}

\maketitle

\section{Introduction}
Graphene has a honeycomb geometry consisting of carbon atoms arranged in the vertices of a regular hexagon. 
It has been considered as a new benchmark material for the construction of next-generation nanoelectronic 
devices due to its unique geometry of the atoms \cite{novoselov11, neto, singh, 
zhu, jiang, lee,mccann}. Over the years, experimental progress in graphene shows that it has very
high electrical and thermal conductivity \cite{xu,ghosh,balandin}, electronic mobility 
\cite{novoselov1,geim} and display very high optical transmittance \cite{kymakis}. Apart from that, 
graphene can be used to fabricate electronic devices like ultrafast transistor \cite{lin,mattevi,jangid}, 
capacitors \cite{liu1,tan,purkait}, electrode \cite{brownson} and for various other sensing devices 
\cite{xu1,wu1,pena}. Although monolayer graphene devices have a lot of potential applications but Bi-Layer 
Graphene (BLG) has superiority in its conductivity \cite{dean}, mechanical strength \cite{zhang1}, electrical mobility \cite{gosling} and chemical stability \cite{tran}. Moreover, the possibility to efficiently tune the electrical properties by changing the carrier density through a grating or doping \cite{novoselov,novoselov2,ohta2} and its ability to be chemically functionalized [29] make it distinctive from others. Thus, BLG has received significant attention in device fabrications and other spintronic applications. In contrast to monolayer graphene, BLG can exist in three geometries: AA stacking, AB (Bernal) stacking, and twisted bilayer \cite{liang,rozhkov,bagchi, song, alavirad,santos, hu}. Although AA-BLG appears in its simplest form yet, this stacking is found to be less stable than the Bernal phase or AB stacking. In AB stacking, half of the carbon atoms of the top layer lie above the carbon atoms of the bottom layers, while the other atoms are present above the centre of the hexagons of the bottom layer. 

The quantum transport in BLG and BLG-based hybrids are currently among the foremost actively investigated topics in graphene physics \cite{novoselov1,mccann2, guinea,ohta31,
oostinga,castro, gorbachev,morozov, feldman,xiao21,koshino21,cserti,cserti2,snyman}. Over the past few decades,  efforts have been made to investigate transmission and conductance in various experimentally realizable multi-layer BLG hybrids, and graphene superlattices \cite{cserti, cserti2, snyman,katsnelson,katsnelson2,katsnelson3,azarova}. Recently, some studies have been undertaken to understand the spin dependent quantum transport in mono and bilayer graphene nanostructures in the presence of Rashba Spin-Orbit Coupling (RSOC) \cite{liu11,zhang111, chico,ganguly1,ganguly2, fouladi1,zhang121,liu111}. RSOC is an asymmetric spin-orbit coupling responsible for splitting of energy subbands \cite{rashba}, observed in metal surfaces \cite{lashell}, interfaces 
\cite{acharjee1,acharjee2,acharjee3,cavigilia} and also in bulk materials \cite{acharjee1,acharjee2,acharjee3,cavigilia,ishizaka}. Moreover, it has given rise to new phenomena such as spin currents and the spin Hall Effect \cite{shinova} as well. 

Recent works indicate that some Rashba materials may also display the Zeeman effect, which is a momentum-independent spin splitting of the energy bands due to the interaction of spin with an external magnetic field or an internal exchange field.
The discovery of the Rashba Zeeman (RZ) effect in Ag$_2$Te/Cr$_2$O$_3$ heterostructure stimulated vigorous research, owing to its potential to change the fundamental properties of a material \cite{tao1}. It provides novel features not found in pure Rashba or Zeeman systems. For example, insulator-to-conductor transition can be triggered by the exchange field via the RZ \cite{tao1} effect. Also, spin-dependent transport properties of a quantum wire in the presence of the RZ effect may lead to the design of a spin filter device \cite{xiao}. It has been demonstrated that the Rashba spin can be altered by using magnetization switching \cite{zhai} while the magnetization can be reversed by the polarization switching \cite{krempasky}. Moreover, the RZ effect also plays an essential role in changing the quantum properties. For example, the conductance plateau in quantum-point-contact InSb nanowires with sizable Rashba Spin-Orbit Coupling could be tuned from $e^2/h$ to $2e^2/h$  by the magnetic field orientation \cite{kammhuber}. The RZ effect also plays a crucial role in the spin magnetization and thus also responsible for spin dependent transport in graphene.

Although significant efforts have been made to understand spin dependent quantum transport in the graphene nanoribbons and AA-BLG considering pure Rashba or Zeeman materials as previously mentioned, but the spin dependent transport in AB-BLG and RZ heterostructure are missing. 
Thus, in this work, we have investigated the spin-dependent quantum transport in RZ$|$BGL$|$RZ hybrid in clean limit. We have studied the role of magnetization strength and orientation, RSOC and the bias voltage in transmission and conductivity of the proposed structure. 

The organization of this paper are as follows: we present a minimal theory to study ballistic transport and calculate the scattering coefficients. The conductance is studied by using Landauer-B\"{u}ttiker formula for RZ$|$BGL$|$RZ hybrid in section II. In Section III, we study the spin dependence, the effect of bias voltage and RSOC on the transmission coefficients, conductivity and Fano factor. Finally, we present a summary of our work in Section IV.

\section{Minimal Theory and Scattering Coefficients}
\begin{figure}[hbt]
\centerline
\centerline{
\hspace{0.5cm}
\includegraphics[scale=0.53]{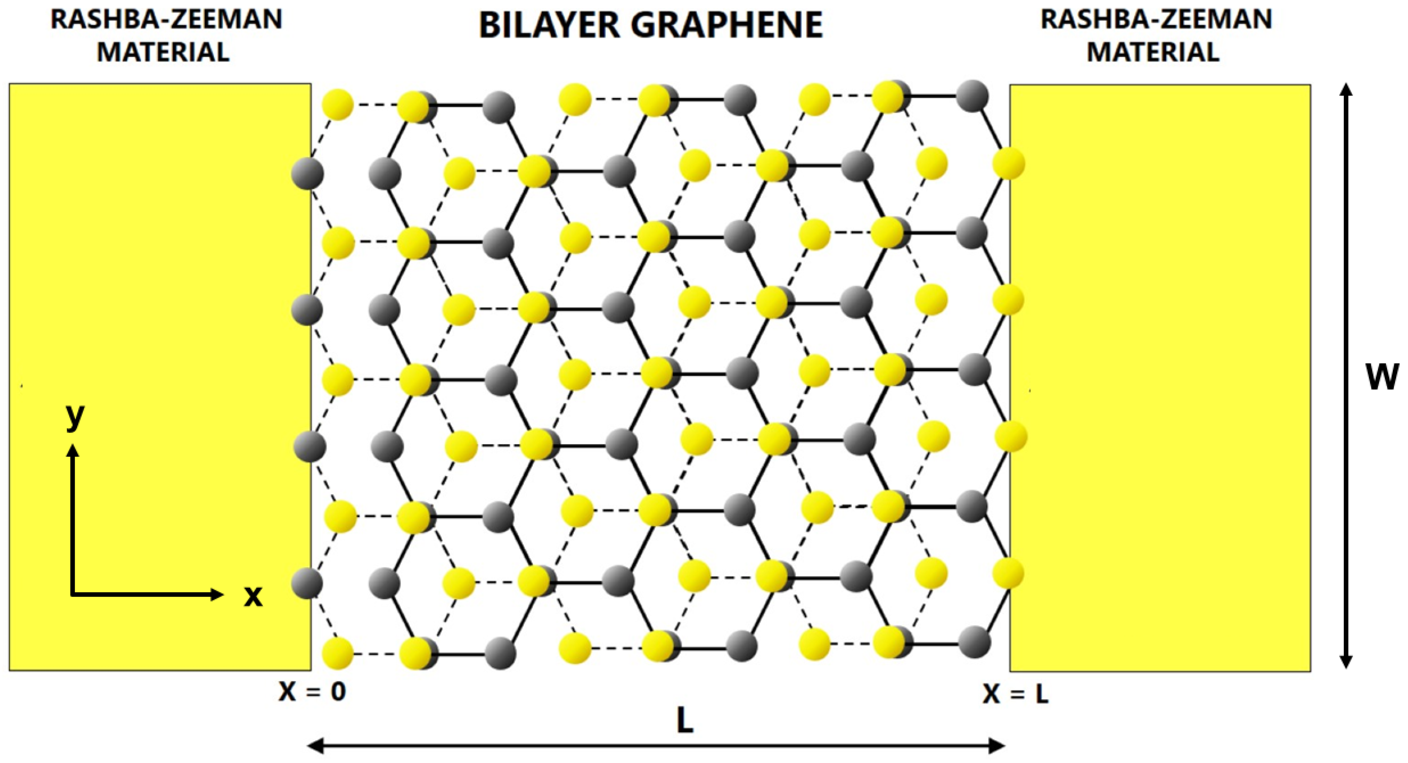}
\vspace{0.1cm}
\includegraphics[scale = 0.36]{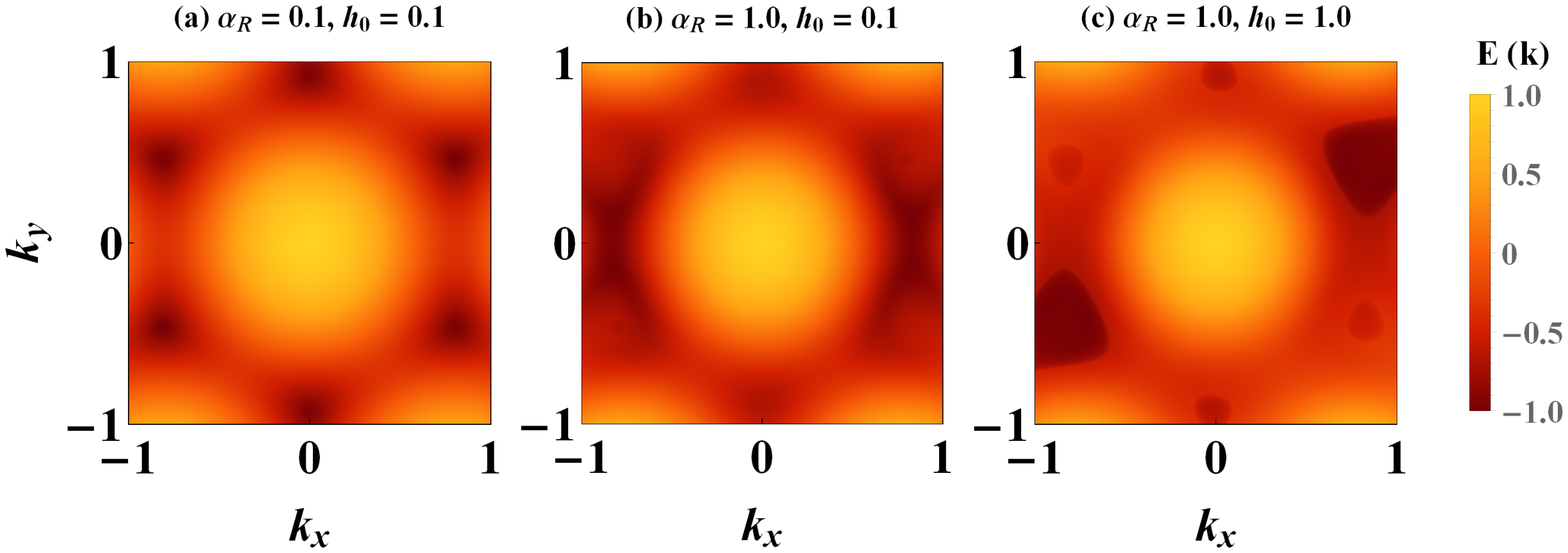}
\hspace{-0.05cm}
\includegraphics[scale = 0.35]{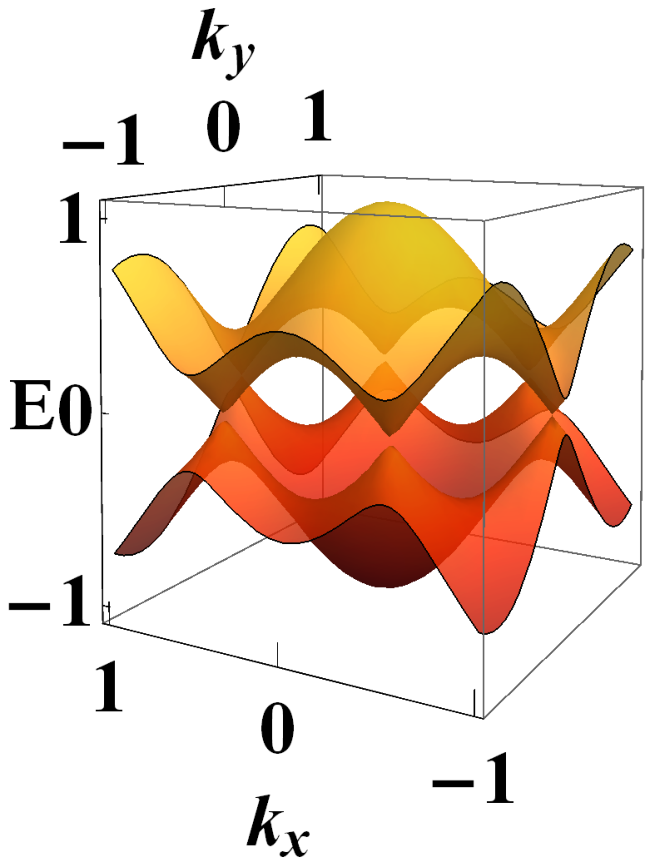}
\hspace{-0.5cm}
\includegraphics[scale = 0.35]{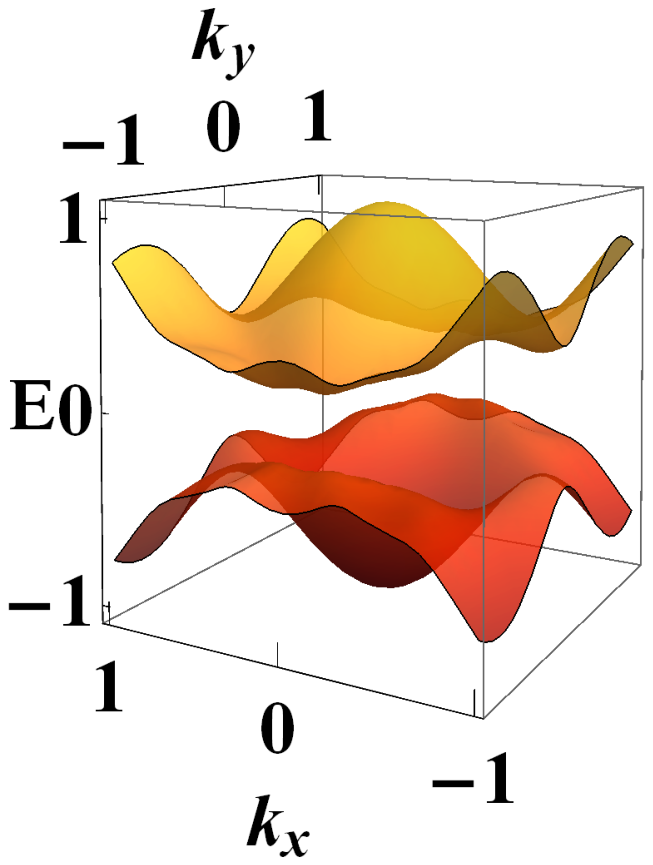}
\hspace{-0.5cm}
\includegraphics[scale = 0.35]{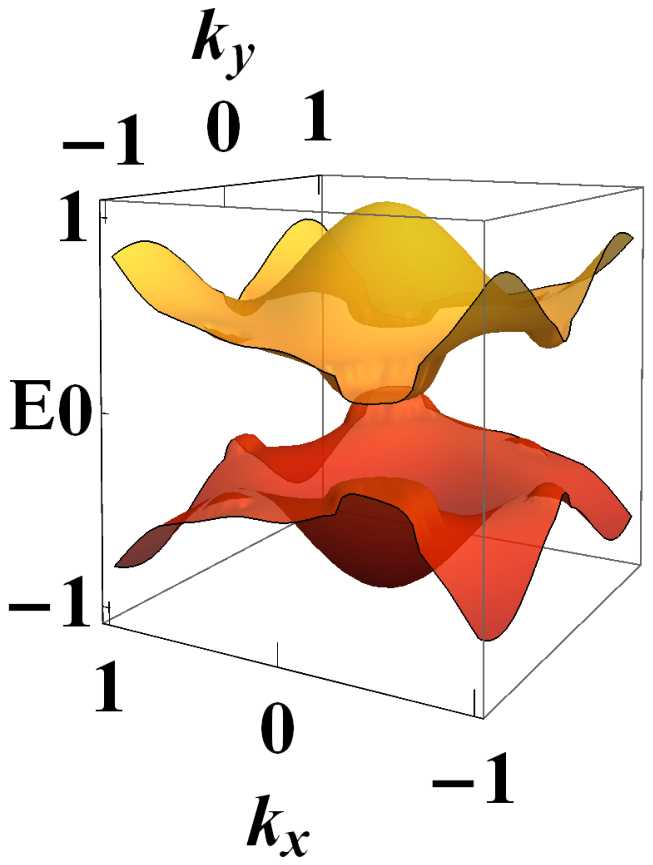}
\hspace{-0.5cm}
}
\caption{(Top panel) Schematic representation of two probes Rashba-Zeeman (RZ) - Bi-Layer Graphene (BLG) hybrid device with armchair edges of length L and zig-zag edges of width W. The yellow-shaded regions correspond to semi-infinite RZ leads. Here we consider AB-stacking with A and B atoms represented by black and yellow circles, respectively. (Middle panel) Density plot to study the variation of energy spectrum of RZ$|$BLG$|$RZ hybrid for different choices of $\alpha_{\text{R}}$ and $h_0$, considering, $\zeta = 0.3\pi$ and $\phi = 0.3\pi$. (Bottom panel) Three-dimensional plot to understand the change in Dirac points of the energy spectrum for different $\alpha_{\text{R}}$ and $h_0$ of RZ$|$BLG$|$RZ hybrid. We consider, $\text{eV} = 0.1$ for plot (a) while $\text{eV} = 1$ for plots (b) and (c).
}
\label{fig1}
\end{figure}

Ballistic transport in finite mesoscopic bilayer and its hybrids has been studied in several works \cite{katsnelson,katsnelson2,katsnelson3,cserti,snyman,cserti2,mccann} in recent times.
In this work, we consider a two probe AB - bilayer graphene (BLG) hybrid with armchair edges of length L and zig-zag edges of width W connected to a semi-infinite RZ leads as shown in the top panel of Fig. \ref{fig1}. We followed the wave-matching approach of Snyman and Beenakker \cite{snyman} to calculate scattering amplitudes. The effective tight-binding Hamiltonian for the proposed RZ$|$BLG$|$RZ system in the presence of an electric field can be written as \cite{mccann,acharjee1,acharjee2,acharjee3, rozhkov1, abdullah, masir}
\begin{multline}
\label{eq1}
\mathcal{H}^{\text{eff}}_k = - \gamma_0 \sum_{j, \sigma} (\hat{a}^\dagger_{j1\sigma}\hat{a}_{j2\sigma}+\hat{b}^\dagger_{j1\sigma}\hat{b}_{j2\sigma})
+\gamma_1 \sum_{j, \sigma}(\hat{a}_{j2\sigma}^\dagger \hat{b}_{j1\sigma} \\ + \hat{b}^\dagger_{j1\sigma}\hat{a}_{j2\sigma})
-\gamma_3 \sum_{j, \sigma} (\hat{a}^\dagger_{j1\sigma}b_{j2\sigma} + \hat{b}^\dagger_{j2\sigma}\hat{a}_{j1\sigma})\\
+\gamma_4\sum_{\alpha,\sigma} (\hat{a}^\dagger_{j\alpha\sigma}\hat{b}_{j\alpha\sigma} + \hat{b}^\dagger_{j\alpha\sigma}\hat{a}_{j\alpha\sigma}) +  \sum_{k, \sigma} \epsilon_k  \hat{a}^\dagger_{k\sigma} \hat{a}_{k\sigma}
 \\- \sum_{k\alpha\beta}a^\dagger_{k\alpha}(\mathbf{\hat{h}} . \boldsymbol{\hat{\sigma}})_{\alpha\beta}\hat{a}_{k\beta} +\sum_{k\alpha\beta}\hat{a}^\dagger_{k\alpha}
(\mathbf{\hat{g}}_k.\boldsymbol{\hat{\sigma}})_{\alpha\beta}\hat{a}_{k\beta}\\
 + \frac{\text{eV}}{2} \sum_{j,\sigma} (\hat{a}^\dagger_{j\alpha\sigma}\hat{a}_{j\alpha\sigma} + \hat{b}^\dagger_{j\alpha\sigma}\hat{b}_{j\alpha\sigma})  + h. c. 
\end{multline}
\noindent where, $\hat{a}^\dagger_{i\alpha\sigma}$ ($\hat{b}^\dagger_{i\alpha\sigma}$) and $\hat{a}_{i\alpha\sigma}$ ($\hat{b}_{i\alpha\sigma}$) are the creation and annihilation operators of the A (B) sublattices, respectively corresponding to the site j = 1, 2 and spin $\sigma$. The parameter $\gamma_0 =  \langle\Phi_{\text{A1}}|\hat{\textsc{H}}|\Phi_{\text{A2}}\rangle = \langle\Phi_{\text{B1}}|\hat{\textsc{H}}|\Phi_{\text{B2}}\rangle$ correspond to hopping energy of A$_1$ (B$_1$) and A$_2$ (B$_1$) atoms while $\gamma_1 = \langle\Phi_{\text{A2}}|\hat{\textsc{H}}|\Phi_{\text{B1}}\rangle$ is the hopping energy between A$_2$ and B$_1$  atoms. Here $\gamma_3 = \langle\Phi_{\text{A1}}|\hat{\textsc{H}}|\Phi_{\text{B2}}\rangle$ is the hopping energy between A$_1$ and B$_2$  atoms while $\gamma_4 = \langle\Phi_{\text{A1}}|\hat{\textsc{H}}|\Phi_{\text{B1}}\rangle = \langle\Phi_{\text{A2}}|\hat{\textsc{H}}|\Phi_{\text{B2}}\rangle$ is the hopping energy between A$_1$ (A$_2$) and B$_1$  (B$_2$) atoms. The term $\text{eV}$ corresponds to applied biasing energy measured in fermi energy (E$_\text{F}$). Here, $\alpha = 1, 2$ corresponding to the $j^{th}$ site while $\epsilon_k$ represents the kinetic energy term. The last two terms of Eq. (\ref{eq1}) corresponds to the RZ effect where $\mathbf{\hat{h}}$ represents the exchange energy and $\mathbf{\hat{g}}_k$ arises due to RSOC while $\boldsymbol{\hat{\sigma}} = (\sigma_x,\sigma_y,\sigma_z)$ are the Pauli's spin matrices. The magnetization vector of RZ material can be defined as 

\begin{align}\notag
\mathbf{\hat{h}} &= (h_x,h_y,h_z) \\\notag
& = (h_0 \sin\zeta\cos\phi, 
h_0 \sin\zeta\sin\phi, h_0 \cos\zeta)
\end{align}

\noindent where $h_0$ is the magnetization strength measured in E$_\text{F}$. The parameters $\zeta$ and $\phi$ are the polar and azimuthal angles of magnetization. The vector function $\mathbf{\hat{g}}_k$ in the last term of Eq. (\ref{eq1}) accounts for the antisymmetric spin-orbit coupling. Introducing four-component spinor basis $\Psi_{k\sigma} = \left[\hat{a}_{k1\sigma}, \hat{a}_{k2\sigma}, \hat{b}_{k1\sigma}, \hat{b}_{k2\sigma}\right]^\text{T}$ and considering the mean-field Hamiltonian $ \hat{\mathcal{H}} = \varepsilon_0 +\frac{1}{2} \sum_k \hat{\Psi}^\dagger_{k\sigma}\mathcal{H}_k \hat{\Psi}_{k\sigma}$
one can obtain the Hamiltonian of RZ$|$BLG$|$RZ  hybrid system as \cite{mccann,acharjee1,acharjee2,acharjee3, rozhkov1, abdullah, masir} 
\begin{widetext}
\begin{equation}
\label{eq2}
\mathcal{H}_k = \left(
\begin{array}{cccc}
\epsilon^+_{\textsc{A1}}+h'_z+\frac{\text{eV}}{2} & -\gamma_0 f(k)+g_{k_-}+h_{xy} & \gamma_4 f(k) & -\gamma_3 f^\ast(k) \\
 -\gamma_0 f^\ast(k)+ g_{k_+}+h^*_{xy} &  \epsilon^+_{\textsc{B1}}-h'_z-\frac{\text{eV}}{2} & \gamma_1 & \gamma_4 f(k) \\
 \gamma_4 f^\ast(k) &  \gamma_1 & \epsilon^-_{\textsc{A2}}-h'_z+\frac{\text{eV}}{2} & -\gamma_0 f(k)+g_{k_+}-h^*_{xy} \\
 -\gamma_3 f(k) &  \gamma_4 f^\ast(k) & -\gamma_0 f^\ast(k)+g_{k_-}-h_{xy} & \epsilon^-_{\textsc{B2}}+h'_z-\frac{\text{eV}}{2} \\
\end{array}
\right)
\end{equation}
\end{widetext}
\noindent where, $\epsilon^\pm_{\text{A}j (\text{B}j)} = \epsilon_{\text{A}j (\text{B}j)} \pm \epsilon_0$ with $j = 1, 2$ for respective sites. We consider, $\epsilon_{\textsc{A1}} = \frac{1}{2}(-\textsc{U} +\delta_{\textsc{AB}})$, $\epsilon_{\textsc{B1}} = \frac{1}{2}(-\textsc{U} +2\Delta'-\delta_{\textsc{AB}})$, $\epsilon_{\textsc{A2}} = \frac{1}{2}(\textsc{U} +2\Delta'+\delta_{\textsc{AB}})$ and  $\epsilon_{\textsc{B2}} = \frac{1}{2}(\textsc{U} -\delta_{\textsc{AB}})$ are the on-site energies corresponding to the atomic sites $\textsc{A1}$, $\textsc{B1}$, $\textsc{A2}$ and $\textsc{B2}$ respectively \cite{mccann}. Here, $\textsc{U}$ is a 
parameter which characterizes the asymmetry between the two layers \cite{mccann2,guinea,ohta31,mccann3,min,
oostinga,castro}, $\Delta'$ is a parameter used to characterize the energy difference between the dimer and the non-dimer sites \cite{nilsson,zhang11, li,dresselhaus} while $\delta_\textsc{AB}$ represents the energy difference between A and B atomic sites \cite{mucha}. The function $f(k)$ is used to characterize the hopping of the BLG \cite{rozhkov}. For the proposed BLG geometry, the Brillouin zones are located at wavevector $K_\xi = \xi(0, 4\pi/3a)$ \cite{mccann,snyman}. Considering only the interlayer asymmetry $\epsilon_0$ in the on-site energy and 
neglecting $\gamma_4$ one can approximate the on-site energies as \cite{mccann}: $\epsilon_\textsc{A1} = \epsilon_\textsc{B1} = -\textsc{U}/2$ and $\epsilon_\textsc{A2} = \epsilon_\textsc{B2} = \textsc{U}/2$.
Moreover, to characterize the interlayer coupling of the BLG, we consider a new length scale $l_1=\hbar v/\gamma_1$. We assume that $L>>l_1$ so that the graphene system can be treated as a bilayer rather than two separate monolayers \cite{mccann}. The parameters $\epsilon_0$, $h'_z$, $h_{xy}$ and $g_{k\pm}$ appeared in Eq. (\ref{eq2}) are defined as, 
\begin{align}
\epsilon_0 &= -\mu_0 + \textsc{U}_0\delta_1,\nonumber \\
h'_z &= h_z\Theta_2,\nonumber\\
h_{xy} &= (h_x-ih_y)\Theta_2,\nonumber \\
g_{k\pm} &= \alpha_{\textsc{R}}(k_x \mp i k_y)[\Theta(-x)+\Theta(x-L)],\nonumber\\
\delta_1 &= [\delta(x) +\delta(x-L)],\nonumber\\
\Theta_1 &= [\Theta(-x) + \Theta(x-L)],\nonumber\\
\Theta_2 &= [\Theta(x)\Theta(L-x)],\nonumber
\end{align}
where, $\mu_0 = \mu_{\textsc{RZ}}\Theta_1 + \mu_{\textsc{BLG}}\Theta_2$ is the chemical potential of the system and $\textsc{U}_0$ is the non-magnetic delta like barrier potential. The parameter $\alpha_\textsc{R}$ accounts for the strength of RSOC while $\delta(x)$ and $\Theta(x)$ respectively are the Dirac delta and Heavy-side step functions.

For an incoming electron incident at an angle $\theta$ at left RZ$|$BLG interface, the trajectory can be expressed in terms of the momentum components  $k_x = |\mathbf{k}|\cos\theta$ and $k_y = |\mathbf{k}|\sin\theta$. The total wave function of the incoming electron can be defined as
\begin{multline}
\label{eq3}
\Psi(x) = \big[\Psi^{\textsc{RZ (L)}}_\pm (x)\Theta(-x)+\Psi^{\textsc{BLG}}_\pm(x)\Theta(x)\Theta(L-x)
\\+\Psi^{\textsc{RZ (R)}}_\pm (x)\Theta(x-L)\big]e^{ik_yy}
\end{multline}
where, $\Psi^{\textsc{RZ (L)}}_\pm (x)$, $\Psi^{\textsc{BLG}}_\pm(x)$, 
$\Psi^{\textsc{RZ (R)}}_\pm (x)$ are the wave functions of the incoming electron in 
the left RZ, BLG and right RZ leads respectively. $k_y$ is the momentum parallel to the interface. 

The wave function of the BLG region can be defined as \cite{snyman,mccann}
\begin{multline}
\label{eq4}
\Psi^{\textsc{BLG}}_\pm (0 < x < L) = c^\pm_1(-i\hat{\tau}_1 - \hat{\tau}_2 + \hat{\tau}_3 - i\hat{\tau}_4) 
\\+ c^\pm_2 (i\hat{\tau}_1 + \hat{\tau}_2 + \hat{\tau}_3 - i\hat{\tau}_4)e^{i\kappa^+ x} 
+ c^\pm_3(i\hat{\tau}_1 - \hat{\tau}_2 + \hat{\tau}_3 + i\hat{\tau}_4) 
\\+ c^\pm_4 (-i\hat{\tau}_1 + \hat{\tau}_2 + \hat{\tau}_3 + i\hat{\tau}_4)e^{i\kappa^- (x-\textsc{L})}
\end{multline}

The wavefunction corresponding to left and right RZ leads can be defined as \cite{linder,acharjee1,acharjee2,acharjee3}
\begin{multline}
\label{eq5}
\Psi^{\textsc{RZ (L)}}_\pm (x < 0)  = s_{+}[\hat{\tau}_1 a + \hat{\tau}_2 b e^{-i\phi}]e^{i k^{+}\cos\theta x}
\\+s_{-}[\hat{\tau}_1 b e^{i\phi} + \hat{\tau}_2 a]e^{ik^{-}\cos\theta x} +r_+^{\pm}[\hat{\tau}_1 a
+ \hat{\tau}_2 b e^{i\phi} 
\\- \hat{\tau}_3 b e^{-i\phi}
+ \hat{\tau}_4 a]e^{\mp i k^{\pm}s_\pm\cos\theta x}
+r_-^{\pm}[- \hat{\tau}_1 b e^{-i\phi} 
\\+ \hat{\tau}_2 a + \hat{\tau}_3 a 
+  \hat{\tau}_4 b e^{i\phi}]e^{\pm i k^{\pm}s_\mp\cos\theta x}
\end{multline}
\begin{multline}
\label{eq6}
\Psi^{\textsc{RZ (R)}}_\pm (x > L)  = t_+^{\pm}[\hat{\tau}_1 a
+ \hat{\tau}_2 b e^{i\phi} - \hat{\tau}_3 b e^{-i\phi}
\\+ \hat{\tau}_4 a]e^{\pm i k^{\mp}\cos\theta (x-\textsc{L})}
+t_-^{\pm}[- \hat{\tau}_1 b e^{-i\phi} 
\\+ \hat{\tau}_2 a + \hat{\tau}_3 a 
+  \hat{\tau}_4 b e^{i\phi}]e^{\mp i k^{\pm}\cos\theta (x-\textsc{L})}
\end{multline}
\noindent where, $r_\pm^{\pm}$ and $t_\pm^{\pm}$ are scattering amplitudes of the reflected and the transmitted wave respectively. We consider, ($s_+ = 1$, $s_- =0$) for incoming up-spin electrons while ($s_+ = 0$, $s_- =1$) for down-spin electron. Here, $\hat{\tau}_1 = (1,0,0,0)^\textsc{T}$, $\hat{\tau}_2 = (0,1,0,0)^\textsc{T}$, $\hat{\tau}_3 = (0,0,1,0)^\textsc{T}$ and $\hat{\tau}_4 = (0,0,0,1)^\textsc{T}$. The parameters $a$ and $b$ appeared in Eqs. (\ref{eq5}) and (\ref{eq6}) can be defined as \cite{linder}
\begin{equation}
\label{eq7}
a = \frac{1}{\sqrt{1 + \left[\frac{h_{xy}}{(h + h_z)}\right]^2}}; \hspace{0.3cm} b = \frac{a h_{xy}}{(h + h_z)} 
\end{equation} 
where $h = (h_x^2+h_y^2+h_z^2)^{\frac{1}{2}}$, $h_{xy} = (h_x^2+h_y^2)^\frac{1}{2}$.   We define the wave vector $|\boldsymbol{\kappa}| \approx \pm(E-\textsc{U})/\hbar v_F$ for the BLG \cite{lejarreta} while the wave vector $k^\pm$ in presence of bias voltage in the RZ-leads can be defined as
\begin{multline}
\label{eq8}
k^\pm_\sigma = \pm\mu_{\textsc{RZ}}\pm[h_y-\alpha_\text{R} k_x)^2+(h_x-\alpha_\text{R} k_y)^2
\\+ (\text{\text{eV}}-2\sigma h_z)^2]^{\frac{1}{2}}
\end{multline}
\noindent where, $\mu_{\textsc{RZ}}$ is the chemical potential of the RZ leads. The parameter $\sigma = \pm 1$, correspond to up and downspin particles respectively. 

The energy band spectra of the RZ$|$BLG$|$RZ hybrid can be studied by diagonalizing the Hamiltonian in Eq. (\ref{eq2}). We study the energy band spectra in the middle and bottom panels of Fig. \ref{fig1} for different $\alpha_{\text{R}}$ and $h_0$ values considering $\zeta = 0.3\pi$ and $\phi = 0.3\pi$.  We have seen a drastic change in the band structure for different choices of $\alpha_{\text{R}}$ and $h_0$. For $\alpha_{\text{R}}  = 0.1$, $h_0 = 0.1$ and $\text{eV} = 0.1$, the band structure is composed of six Dirac points as seen from the plot in Fig. \ref{fig1}(a) which is in accordance with the band structure of pure BLG system \cite{mccann}. The band edges gets flattened along $k_y$ direction for $\alpha_{\text{R}}  = 1.0$, $h_0 = 0.1$ and  $\text{eV} = 1$, which is due to the dominance of Rashba effect. However, in the presence of Rashba-Zeeman interaction, the band edges get flattened and shift nearer to two opposite K-points. Thus two broad dark bands appear in the band spectra for $\alpha_{\text{R}}  = 1.0$, $h_0 = 1.0$ and  $\text{eV} = 1$, as seen from the plot in Fig. \ref{fig1}(c). It is to be noted that in both the plots (b) and (c) of Fig. \ref{fig1}, the bandgap gets widened due to the presence of applied bias voltage $\text{eV} = 1$. 

\subsection{Scattering Amplitudes}
It must be emphasized that there exist eight unknown scattering amplitudes $r_\pm^\pm$, $c_1^\pm$, $c_2^\pm$, $c_3^\pm$, $c_4^\pm$ and $t_\pm^\pm$ for each incoming mode from the left RZ-material. These scattering amplitudes can be obtained by continuity of the wave functions at $x = 0$ and $x = \textsc{L}$.
\begin{equation}
\label{eq9}
\Psi^{\textsc{RZ (L)}}_\pm (0) = \Psi^{\textsc{BLG}}_\pm(0); \hspace{0.3cm}
\Psi^{\textsc{BLG}}_\pm(\textsc{L}) = \Psi^{\textsc{RZ (R)}}_\pm (\textsc{L})
\end{equation}

Using Eq. (\ref{eq9}) and solving for scattering amplitude yields the transmission matrix \cite{masir,snyman}
\begin{equation}
\label{eq10}
\boldsymbol{t}(\text{eV},\theta, \zeta, \phi) = \left(
\begin{array}{cc}
t_+^+ & t_+^- \\
t_-^+ & t_-^-
\end{array}
\right)
\end{equation}
The transmission probability can be obtained by calculating the eigenvalues of the matrix $(\mathbf{t}\mathbf{t}^\dagger)$ \cite{mccann}.  Thus the transmission coefficient $T_\pm$ using Eq. (\ref{eq10}) are
\begin{align}
\label{eq11}
T_\pm &= tr(\boldsymbol{t}\boldsymbol{t}^\dagger) \nonumber \\
&= \frac{\Gamma_1}{\beta_1^2\beta_2^2\Delta^2}\left[\beta_1^2\Gamma_1+2\beta_1\beta_2\Gamma_2+\beta_2^2\Gamma_1
\pm(\beta_1+\beta_2)\Gamma_3\right]
\end{align}
where, $\Gamma_1$, $\Gamma_2$ and $\Gamma_3$ are defined as  
\begin{align}
\Gamma_1 &= 2ab\eta_1\eta_2|\boldsymbol{\kappa}|\sin\phi(1-|\boldsymbol{\kappa}|^2)a^2-b^2\cos2\phi) \nonumber\\
\Gamma_2 &= 2b^2\eta_1\eta_2|\boldsymbol{\kappa}|\sin2\phi(a^2|\boldsymbol{\kappa}|^2+b^2\cos2\phi) \nonumber\\
\Gamma_3 &= \left[\Gamma_1\lbrace\beta_1\Gamma_1(\beta_1-2\beta_2)+\beta_2(4\beta_1 \Gamma_2+\beta_2\Gamma_1)\rbrace\right]^{\frac{1}{2}} \nonumber
\end{align}
\begin{multline}
\label{eq12}
\Delta =(\eta_1-\eta_2)[2a^2 b^2|\boldsymbol{\kappa}|^2+b^2|\boldsymbol{\kappa}|^2(a^2
+b^2|\boldsymbol{\kappa}|^2)
+a^2(2 a^2\\+b^2|\boldsymbol{\kappa}|^2)
+a^2b^2\cos2\phi(1-|\boldsymbol{\kappa}|^2)^2+b^4|\boldsymbol{\kappa}|^2\cos4\phi]
\end{multline}
where, $\beta_1=s_{\pm}e^{\pm ik_+^{\pm }\cos(\theta)\text{L}}$, $\beta_2=s_{\pm }e^{\mp ik_-^{\pm}\cos (\theta)\text{L}}$, $\eta_1=e^{\pm i\kappa^{-}\text{L}}$ and $\eta_2 = e^{\pm i\kappa^{+}\text{L}}$. 
\subsection{Conductance and Shot Noise}
Using the transmission probabilities $T_\pm$ from Eq.(\ref{eq11}), one can find the conductance $G$ and Fano factor $F$ of two terminal RZ$|$BLG$|$RZ system using Landauer-B\"{u}ttiker formulas \cite{buttiker, azarova}. Considering a delta like barrier and within linear bias voltage $G$ and $F$ can be defined as 
\begin{align}
\label{eq13}
G &= G_0 \int_{-\frac{\pi}{2}}^{\frac{\pi}{2}} T_\pm \cos\theta d\theta\\
\label{eq14}
F &= \frac{\int_{-\frac{\pi}{2}}^{\frac{\pi}{2}}T_\pm (1-T_\pm)\cos\theta d\theta}{\int_{-\frac{\pi}{2}}^{\frac{\pi}{2}}T_\pm\cos\theta d\theta}
\end{align}
where, $G_0 = 2ge^2EW/h^2v_F$ with $g$ equal to 4 due to two fold spin and valley degeneracy \cite{azarova}.

\begin{figure*}[hbt]
\centerline
\centerline{
\includegraphics[scale = 0.55]{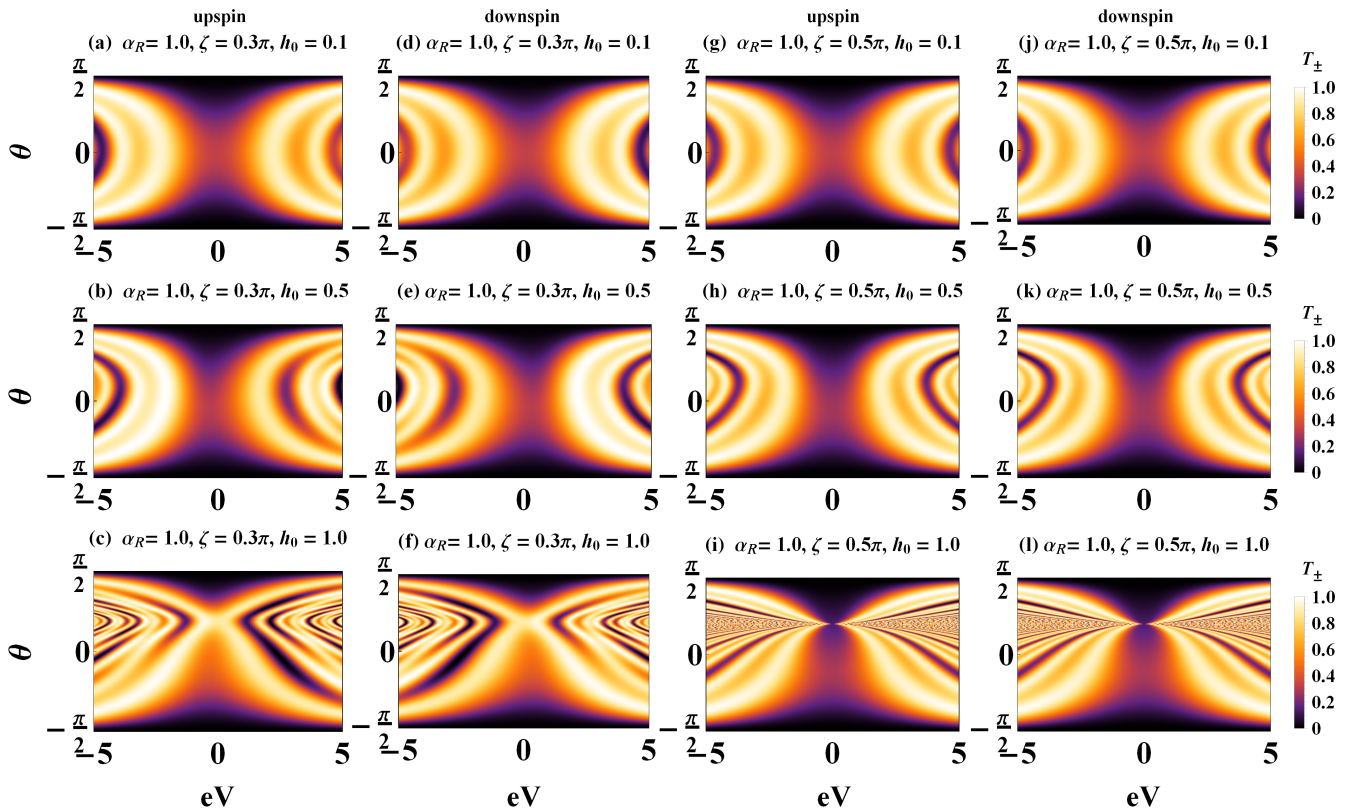}
\hspace{-0.09cm}
}
\caption{Density plot of transmission (T$_\pm$) with $\text{eV}$ and $\theta$ for a double potential barrier in RZ$|$BLG$|$RZ heterostructure.  The plots (a)-(f) are for $\zeta = 0.3\pi$ and $\alpha_{\text{R}} = 1$ while the plots (g)-(l) are for $\zeta = 0.3\pi$ and $\alpha_{\text{R}} = 1$ for different choices of $h_0$.}
\label{fig2}
\end{figure*}

\section{Results and Discussions}
\subsection{Transmission}
\begin{figure*}[hbt]
\centerline
\centerline{
\includegraphics[scale = 0.55]{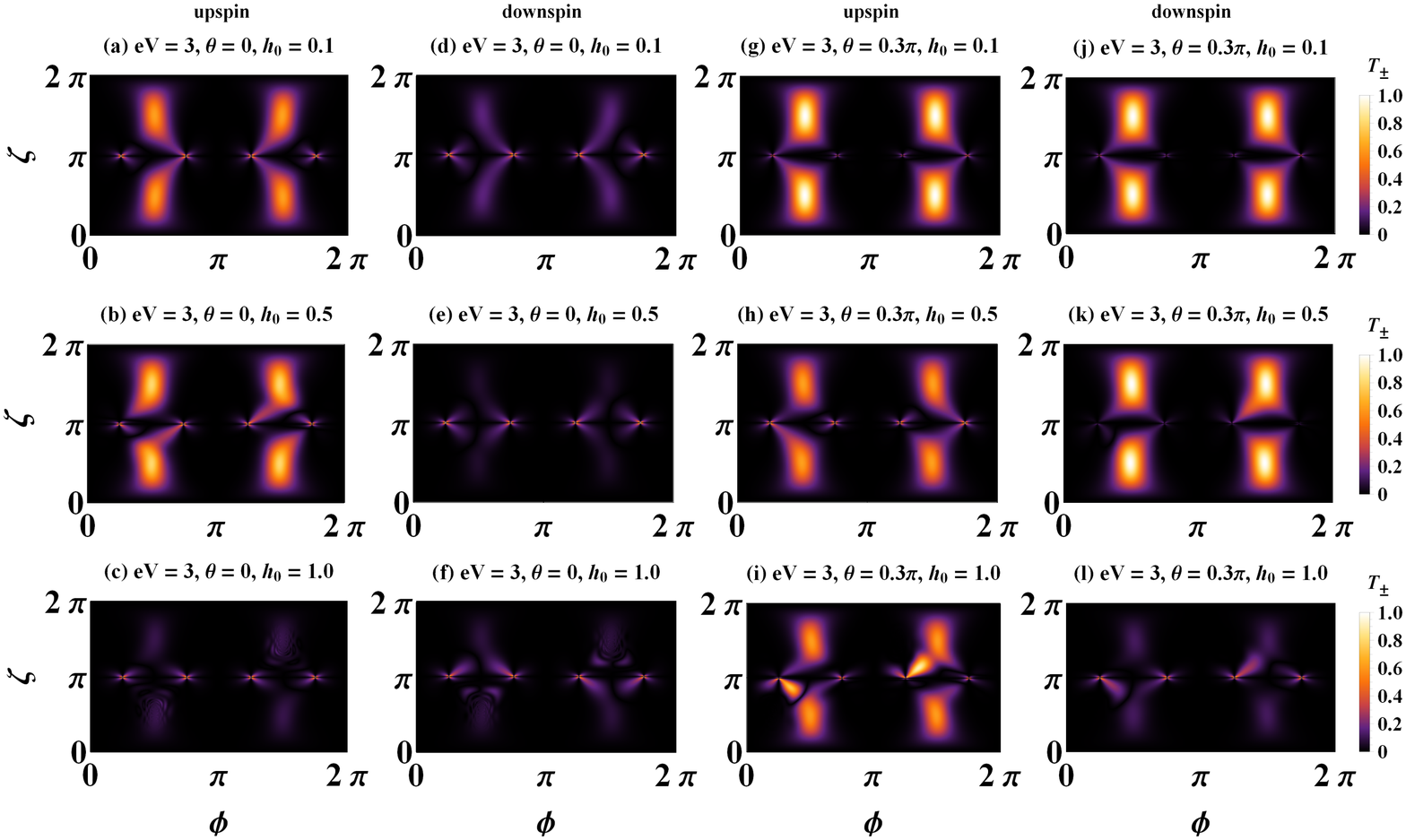}
\hspace{-0.1cm}
}
\caption{Density plot of transmission (T$_\pm$) with $\phi$ and $\zeta$ for a double potential barrier in RZ$|$BLG$|$RZ heterostructure. The plots are for $h_0 = 0.1$ (top), $h_0 = 0.5 $ (middle) and $ h_0 = 1.0$ (bottom) considering $\alpha_{\text{R}} = 1$ and $\text{eV} = 3$. The plots (a-f) in the left panel are for $\theta = 0$  while the plots (h-l) in the right panel are for $\theta = 0.3\pi$.}
\label{fig3}
\end{figure*}
We study the transmission $T_\pm$ for both up and down spin configuration with the angle of incidence $\theta$ and biasing energy $\text{eV}$ in Fig. \ref{fig2}. We set $\alpha_{\text{R}} = 1$, $\phi = 0.3\pi$ and consider different magnetization strength $h_0$ and orientation $\zeta$. Furthermore we choose, $L = 0.1$ and $W = 0.1$ for all our analysis.  We observe that the transmission for up and down spin incoming electrons are quite similar for low values of $h_0$. Some notable points from the transmission spectra in this regime are: (i) The tunnelling is least favourable near the zero bias and in the vicinity of the biasing energy $\text{eV} \approx \pm 5$ for normal incidence condition. But found to be most suitable for moderate and high biasing energies in oblique incidence conditions as seen from Figs. \ref{fig2}(a) and (d). (ii) The transmission is noted to be symmetric for both the angular range of incidence and biasing energy, following the results obtained for tunnelling through a magnetic barrier in BLG \cite{masir}. (iii) There exist two sets of resonance for $h_0 = 0.1$ and $\zeta = 0.3\pi$. A similar characteristics of $T_\pm$ is also observed for $\zeta = 0.5\pi$ and $h_0 = 0.1$ in Figs. \ref{fig2}(g) and (j). Moreover, the transmission is independent of $\zeta$ for low values of $h_0$. 

We observe a drastic change in transmission with the increase in $h_0$. In this condition the number of resonances increased to three for moderate magnetization strength ($h_0 = 0.5$). Also, small decrease in band gap is noticed in this regime for both up and down spin configuration seen from Figs. \ref{fig2}(b) and (e). It is noted that the transmission for up spin electrons are quite different from that of down spin electrons at $h_0 = 0.5$. However, for $\zeta = 0.5\pi$ with $h_0 = 0.5$, the tunnelling for both up and down spins are equally likely seen from Figs. \ref{fig2}(h) and (k). Moreover, the transmission is asymmetric about the angular range in this condition. A significant difference in transmission of both up and down spin configuration are observed for $h_0 \rightarrow \text{E}_{\text{F}}$. In this regime, the number of resonances significantly increased and $T_\pm$ is highly asymmetric about $\theta$. It is to be noted that the angular range of transmission remains same while the energy gap becomes narrow and energy bands overlap for $\zeta = 0.3\pi$ as seen from Figs. \ref{fig2}(c) and (f). It may be due to Klein like tunnelling in graphene heterostructure \cite{masir}.
The resonances becomes too frequent in both up and down spin transmission around the angular range $0$ - $\pi/2$ for $\zeta = 0.5\pi$ with $h_0 = 1$. Moreover, the transmission is asymmetric about the incidence angle and there exists an approximate point contact in the band edges around $\theta = 0.6\pi$ near zero bias condition as seen from Figs. \ref{fig2}(i) and (l). So, it can be concluded that a significant transmission dependence on magnetization orientation exists for strong magnetization strength. 

\subsection{Magnetization dependence of tunnelling}
\begin{figure*}[hbt]
\centerline
\centerline{
\includegraphics[scale = 0.55]{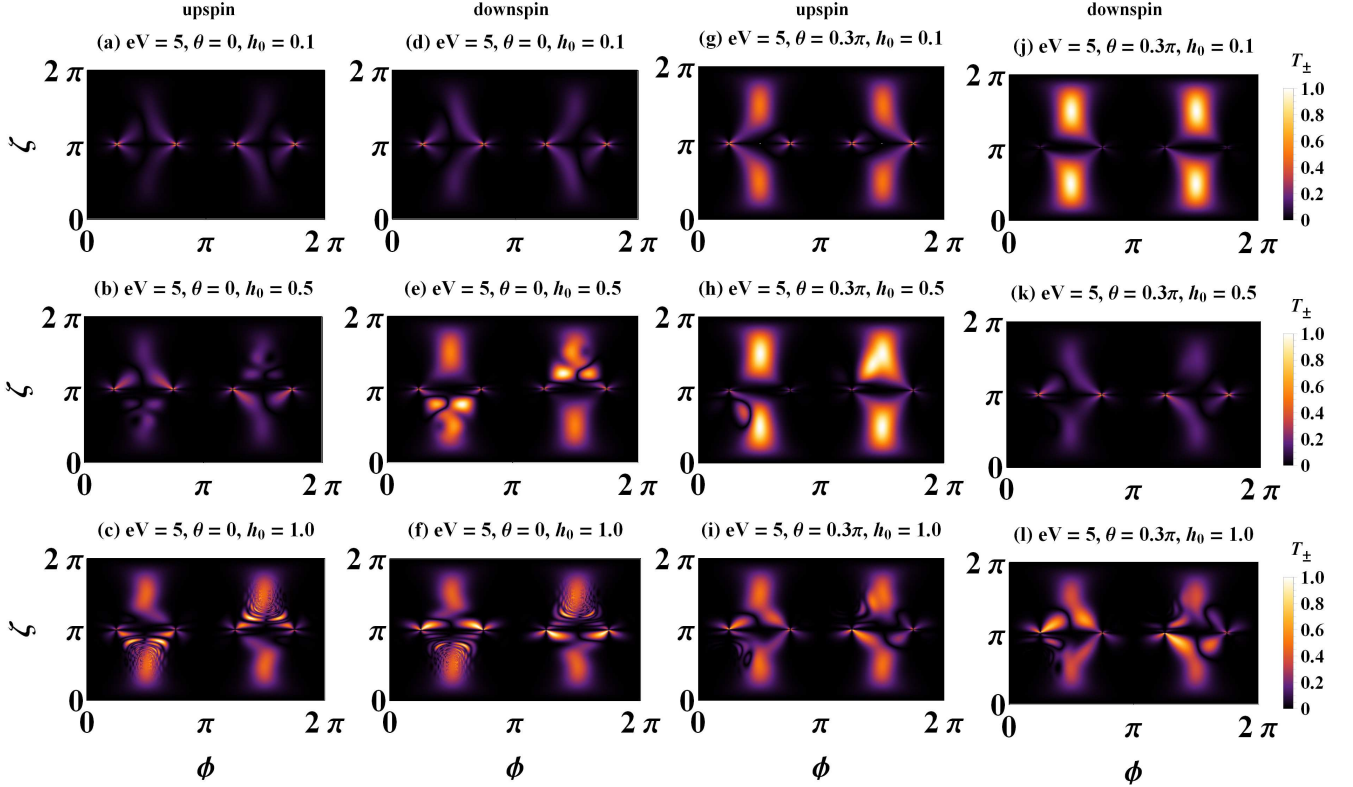}
\hspace{-0.1cm}
}
\caption{Density plot of transmission (T$_\pm$) with $\phi$ and $\zeta$ for a double potential barrier in RZ$|$BLG$|$RZ heterostructure for $h_0 = 0.1$ (top), $h_0 = 0.5 $ (middle) and $ h_0 = 1.0$ (bottom) considering $\alpha_{\text{R}} = 1$ and $\text{eV} = 5$. The plots (a-f) in the left panel are for $\theta = 0$  while the plots (h-l) in the right panel are for $\theta = 0.3\pi$.}
\label{fig4}
\end{figure*}

It is of our interest to understand the magnetization orientation dependence on tunnelling. So, in Figs. \ref{fig3} and \ref{fig4}, we study the transmission with $\zeta$ and $\phi$ for up spin and down spin electrons considering the biasing energies $\text{eV} = 3$ and $\text{eV} = 5$ respectively. We set $\alpha_{\text{R}} = 1$ for this analysis. In plots (a)-(f) of Fig. \ref{fig3}, we consider normal incidence but different $h_0$ values. There exist four sharp maxima around the orientations $(\zeta, \phi) = (\pi, 0.25\pi)$, $(\pi, 0.75\pi)$, $(\pi, 1.25\pi)$ and $(\pi, 1.75\pi)$ while four broad maxima around $\zeta = \pi/2$ and $3\pi/2$. It is to be noted that the sharp maxima are less intense than the broad maxima for up spin incidence at $h_0 = 1$ in both normal and oblique incidence as seen from the plots \ref{fig3}(a) and \ref{fig3}(g). It is to be noted that the tunnelling of up spin at normal incidence is less than that at oblique incidence.  However, we observe an opposite characteristics from plots \ref{fig3}(g) and \ref{fig3}(j) corresponding to $\theta = 0$ and $0.3\pi$ for down spin electrons. In this case, the sharp maxima are more intense than the broad maxima for $\theta = 0$ while the intensity of the broad maxima increases in oblique incidence condition. Similar behaviour is also found for $h_0 = 0.5$ in both up and down spin configurations. In this condition, the intensity of the broad maxima increases, making them more favourable orientations for transmission in normal incidence while the transmission reduces for oblique incidence, noted from Figs. \ref{fig3}(b) and \ref{fig3}(h). Moreover, we also observe that the tunnelling of the down spin electrons for $h_0 = 0.5$ remains quite similar with $h_0 = 0.1$ from Figs. \ref{fig3}(e) and \ref{fig3}(k).  So, even at moderate magnetization ($h_0 = 0.5$) the sharp maxima completely dominates over the broad maxima suggesting that the electrons with the magnetization orientations corresponding to the sharp maxima will be able to tunnel through the system. 

The characteristics of the tunnelling are significantly different for both up and down spin configuration as soon as $h_0$ approaches $\text{E}_{\text{F}}$. For up spin configuration, the orientations corresponding to the sharp maxima dominate over the broad maxima orientations for both $\theta = 0$ and $0.3\pi$. However, the intensities of the broad maxima are more at $\theta = 0.3\pi$ than that at $\theta = 0$. Also, it is noted that the broad maxima regions split into some patches for $h_0 = 1$ as seen from Figs. \ref{fig3}(c) and \ref{fig3}(f). The behaviour of tunnelling of down spin electrons is also different for high magnetization strength. The intensity of broad maxima significantly reduced for $\theta = 0$ and $0.3\pi$. In this case, the four sharp maxima dominate over the four broad maxima. Moreover, the first and third sharp maxima are more intense than the second and fourth maxima, as seen from Figs. \ref{fig3}(i) and \ref{fig3}(l). So, it can be concluded that at high magnetization, the transmission of down spin electrons corresponding to the orientations given by the four sharp maxima is more suitable for tunnelling than the orientations corresponding to four broad maxima. In contrast, for up spin electrons, the orientations corresponding to broad maxima are also preferable for tunnelling at oblique incidence.
\begin{figure*}[hbt]
\centerline
\centerline{
\includegraphics[scale = 0.55]{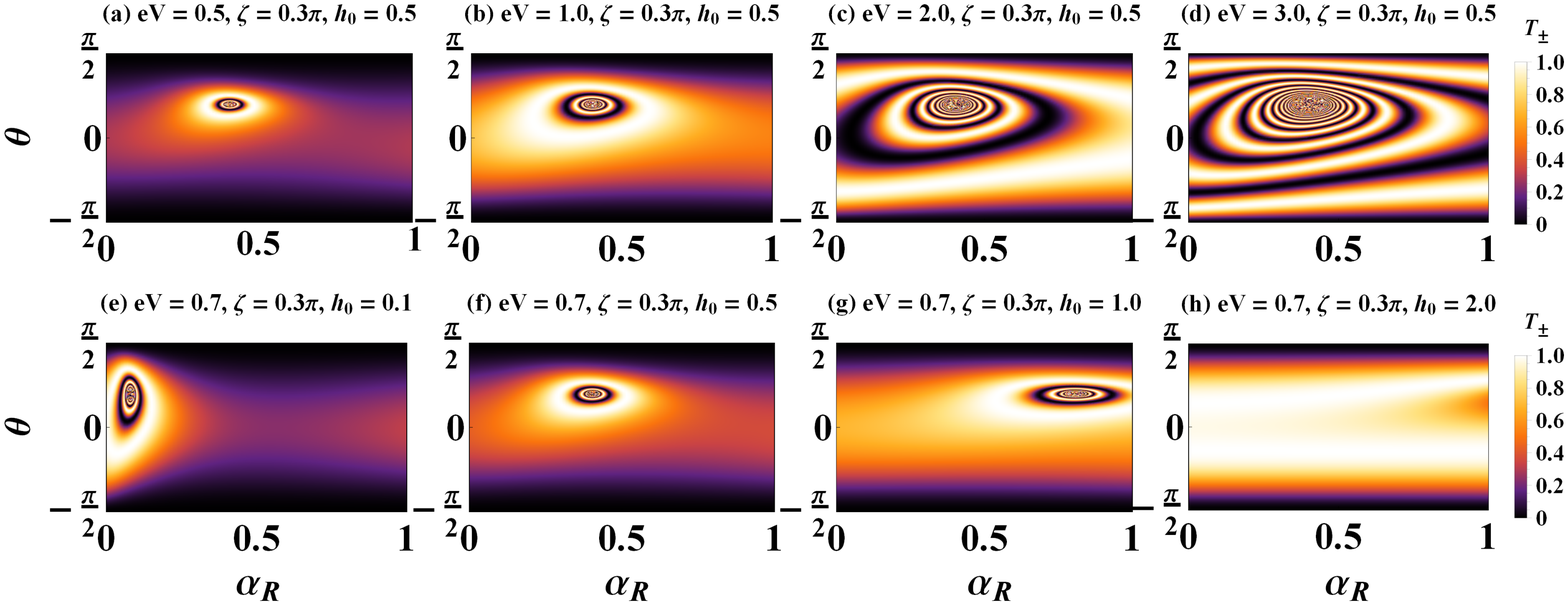}
}
\caption{Density plot of transmission (T$_\pm$) with $\alpha_{\text{R}}$ and $\theta$ for a double potential barrier in RZ$|$BLG$|$RZ structure. The plots in the top panel are for (a) $\text{eV} = 0.5$,  (b) $\text{eV} = 1.0$, (c) $ \text{eV} = 2.0$ and (d) $\text{eV} = 3.0$, considering $h_0 = 0.5$, $\zeta = 0.3\pi$ while the plots in the bottom panel are for (e) $h_0 = 0.1$, (f) $h_0 = 0.5$, (g) $h_0 = 1.0$ and (h) $h_0 = 2.0$ considering $\text{eV} = 0.7$ with $\phi = 0.3\pi$ and $\zeta = 0.3\pi$.}
\label{fig5}
\end{figure*}
From Figs. \ref{fig4}(a) and \ref{fig4}(d), we observe that the characteristics of transmission for up and down spin electrons at normal incidence are similar for $\text{eV} = 5$ and $h_0 = 0.1$. In this condition the orientations $(\pi, 0.25\pi)$, $(\pi, 0.75\pi)$, $(\pi, 1.25\pi)$ and $(\pi, 1.75\pi)$ are found to be most suitable for tunnelling while moderate tunnelling is also possible for orientations $\zeta = \pi/2$ and $3\pi/2$. The system display quite similar behaviour to $\text{eV} = 3$ for oblique incidence condition as seen from plots (g) and (j) of Figs. \ref{fig3} and \ref{fig4}. But for $\text{eV} = 5$ at oblique incidence, the tunnelling probability for up spin electrons with orientations with $\zeta  = \pi/2$ and $3\pi/2$ which corresponds to broad maxima reduces significantly. The broad maxima breaks into some patches of discrete maxima for both up and down spin configurations for $h_0 = 0.5$. It indicates that the magnetization orientation corresponding to the patches of maxima are favoured for tunnelling at normal incidence condition, as seen from Figs. \ref{fig4}(b) and \ref{fig4}(e). However, for oblique incidence $\theta = 0.3\pi$, the orientations corresponding to broader maxima are quite suitable for tunnelling in up spin configuration, while the orientation corresponding to the sharp maxima are favourable for tunnelling seen from Figs. \ref{fig4}(h) and \ref{fig4}(k).

We observe a very different behaviour in Figs. \ref{fig4}(c), \ref{fig4}(f), \ref{fig4}(i) and \ref{fig4}(l) as $h_0$ approaches to $\text{E}_\text{F}$. In this condition, the spin orientation corresponding to both broad and sharp maxima are equally favoured. It is to noted that two broad maxima are continuous while the other two breaks into numerous small individual patches of maxima in Figs. \ref{fig4}(c) and \ref{fig4}(f). It suggests that the transmission is highly sensitive to the spin orientation of the incoming electron at normal incidence conditions for RZ$|$BLG$|$RZ system having magnetization strength of the order of $\text{E}_\text{F}$. Moreover, the system's behaviour is similar even at oblique incidence conditions from Figs. \ref{fig4}(i) and \ref{fig4}(l). It suggests that for high biasing energy and magnetization strength of $\text{E}_\text{F}$, the tunnelling of the electron is independent of its spin and angle of incidence. 

\subsection{RSOC dependence of tunnelling}
Although it is observed from Figs. \ref{fig3} and \ref{fig4} that the tunnelling of electrons in the RZ$|$BLG$|$RZ system is sensitive to magnetization, but the effect of RSOC on transmission is yet to be understood. So, we study the transmission of up and down spin electrons with $\alpha_\text{R}$
and $\theta$ in Fig. \ref{fig5} for different bias voltage $\text{eV}$ and $h_0$ considering $\zeta = 0.3\pi$ and $\phi = 0.3\pi$. We choose different biasing energy $\text{eV}$ with $h_0 = 0.5$ in plots \ref{fig5}(a)-(d). For low $\text{eV}$, we observe that the transmission gradually increases and display
an oscillatory characteristics around $\alpha_\text{R} = 0.45$. As seen from Figs \ref{fig5}(a)-(d), the area of the oscillatory region increases with the increase in $\text{eV}$. However, for all biasing energy, the oscillatory region appears near about $\alpha_\text{R} = 0.45$. Also for moderate biasing, $\text{eV} = 1$, the transmission is possible for all angle of incidence in the vicinity of $-\pi/2$ and $\pi/2$ even when $\alpha_\text{R} \rightarrow 0$ as seen from Fig. \ref{fig5}(b). The area of the oscillatory region grows significantly, indicating the transmission most suitable for certain $\alpha_\text{R}$ values for high biasing, i.e., $\text{eV} = 2$ and $3$, as seen from Figs. \ref{fig5}(c) and (d) respectively. 

To understand the interplay of Zeeman effect with RSOC and their role in tunnelling we have studied the transmission for different $h_0$ values in the plots \ref{fig5}(e)-(h) considering $\text{eV} = 0.7$, $\zeta = 0.3\pi$ and $\phi = 0.3\pi$. It is noted that the appearance of the oscillatory region shifts towards higher RSOC values with the rise in $h_0$. For $h_0 = 0.1,$ the maximum tunnelling probability is found near low $\alpha_\text{R}$ region. It indicates that the tunnelling can be achieved by applying a very low biasing energy in RZ$|$BLG$|$RZ hybrids and considering an RZ material with low RSOC strength and low magnetization strength. With the increase in $h_0$ to $0.5$ in Fig. \ref{fig5}(f), the region of oscillation shifted towards $\alpha_\text{R} = 0.45$ which is similar to Fig. \ref{fig5}(b). The tunnelling can be readily achieved near the angle of incidence $\theta = -\pi/2$ and $\pi/2$ even for materials with low RSOC as soon as $h_0 \rightarrow 1$. The oscillatory region in this condition appear about $\alpha_\text{R} = 0.8$ as seen from Fig. \ref{fig5}(g). The further increase in $h_0$ results in the disappearance of the oscillatory region. However, in this condition, the tunnelling is most favourable in the vicinity of the incident angle $-\pi/2$ and $\pi/2$ as seen from Fig. \ref{fig5}(h). Moreover, the region for the angle of incidence for favourable transmission is enhanced with the increase in $h_0$. It is because the electrons acquire the necessary energy for tunnelling from the Zeeman field even at low biasing conditions. This result signifies that the tunnelling depends on the biasing energy and can be tuned using an RZ material with suitable magnetization strength and RSOC, which follows recent work in BLG with RSOC \cite{zhang22}. 

\begin{figure}[hbt]
\centerline
\centerline{
\includegraphics[scale = 0.41]{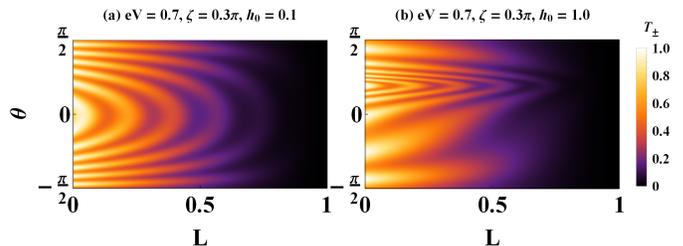}
}
\caption{Density plot of transmission (T$_\pm$) with $L$ and $\theta$ for a double potential barrier in RZ$|$BLG$|$RZ structure. The plots are for (a) $h_0 = 0.5$,  (b) $h_0 = 1.5$ and (c) $ h_0 = 2.0$ considering $\zeta = 0.3\pi$, $\phi = 0.3\pi$ and $\alpha_{\text{R}} = 1$. }
\label{fig6}
\end{figure}

We also investigate the effect of the armchair dimension of the BLG on tunnelling in Fig. \ref{eq6} for different values of $h_0$ considering $\alpha_\text{R} = 1$, $\zeta = 0.3\pi$ and $\phi = 0.3\pi$. For $h_0 = 0.1$, we see that the probability of transmission gradually decreases with length $L$ as seen from Fig. \ref{fig6}(a). However, an oscillatory decrease is observed with length and angle of incidence. The transmission is symmetric about the angular range in this condition. With the rise in $h_0 = 1.0$, the transmission is found to be asymmetric about the angular range. However, it displays oscillatory decay, as seen from Fig. \ref{fig6}(b), which is a very peculiar and anomalous characteristic.

\subsection{Conductance and Fano factor}
\begin{figure*}[hbt]
\centerline
\centerline{
\includegraphics[scale = 0.35]{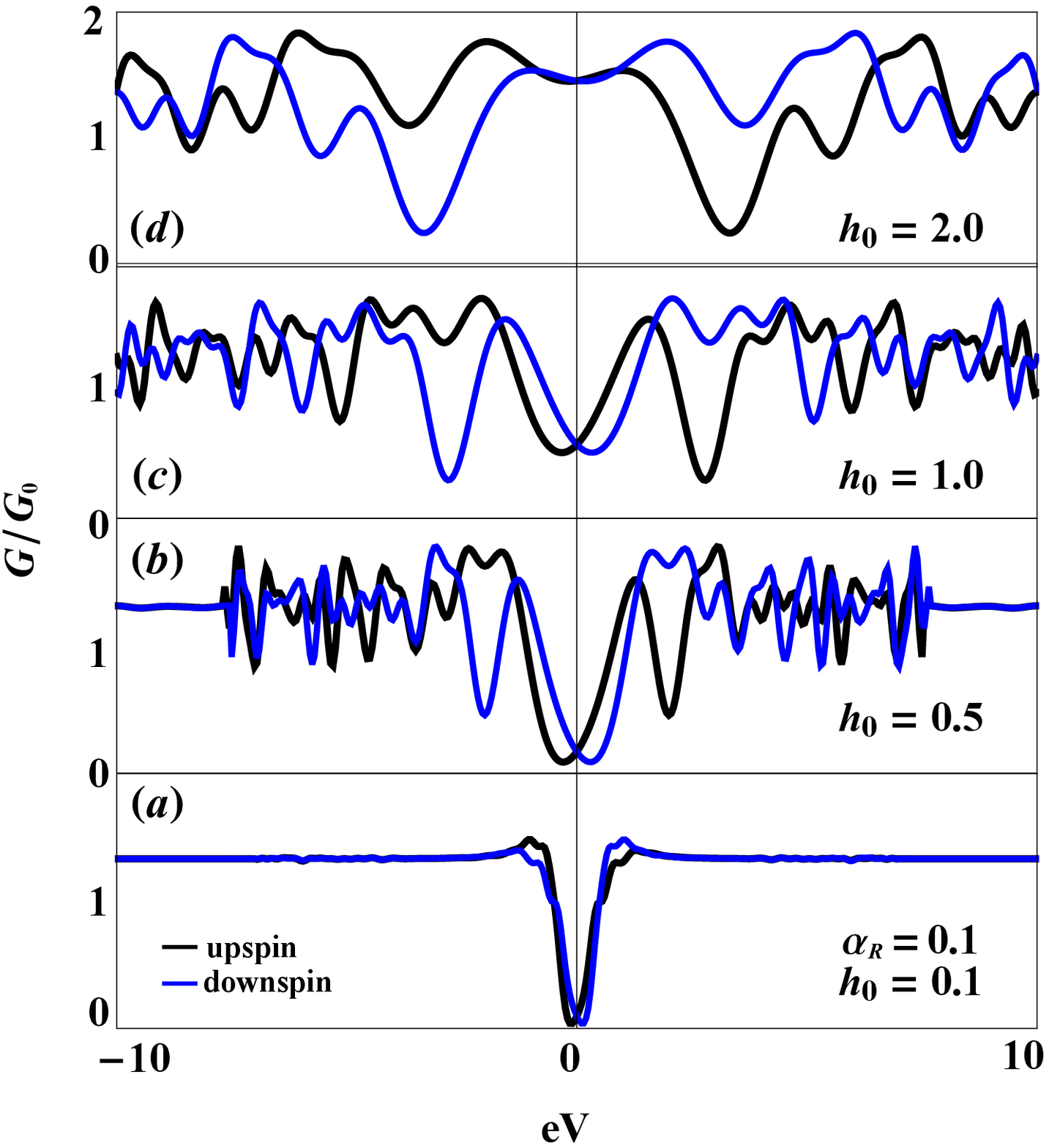}
\hspace{0.3cm}
\includegraphics[scale = 0.35]{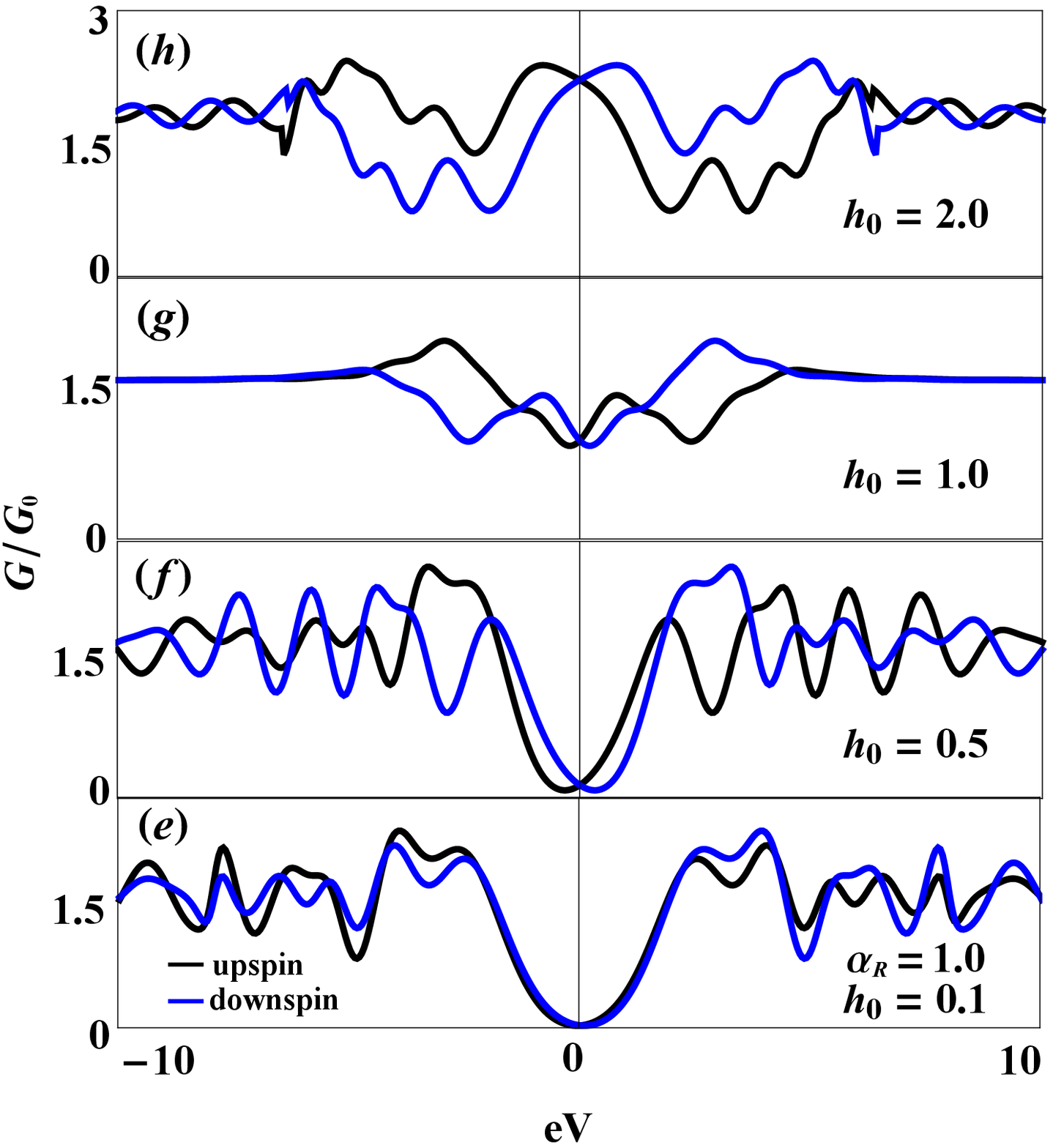}
\hspace{0.3cm}
\includegraphics[scale = 0.35]{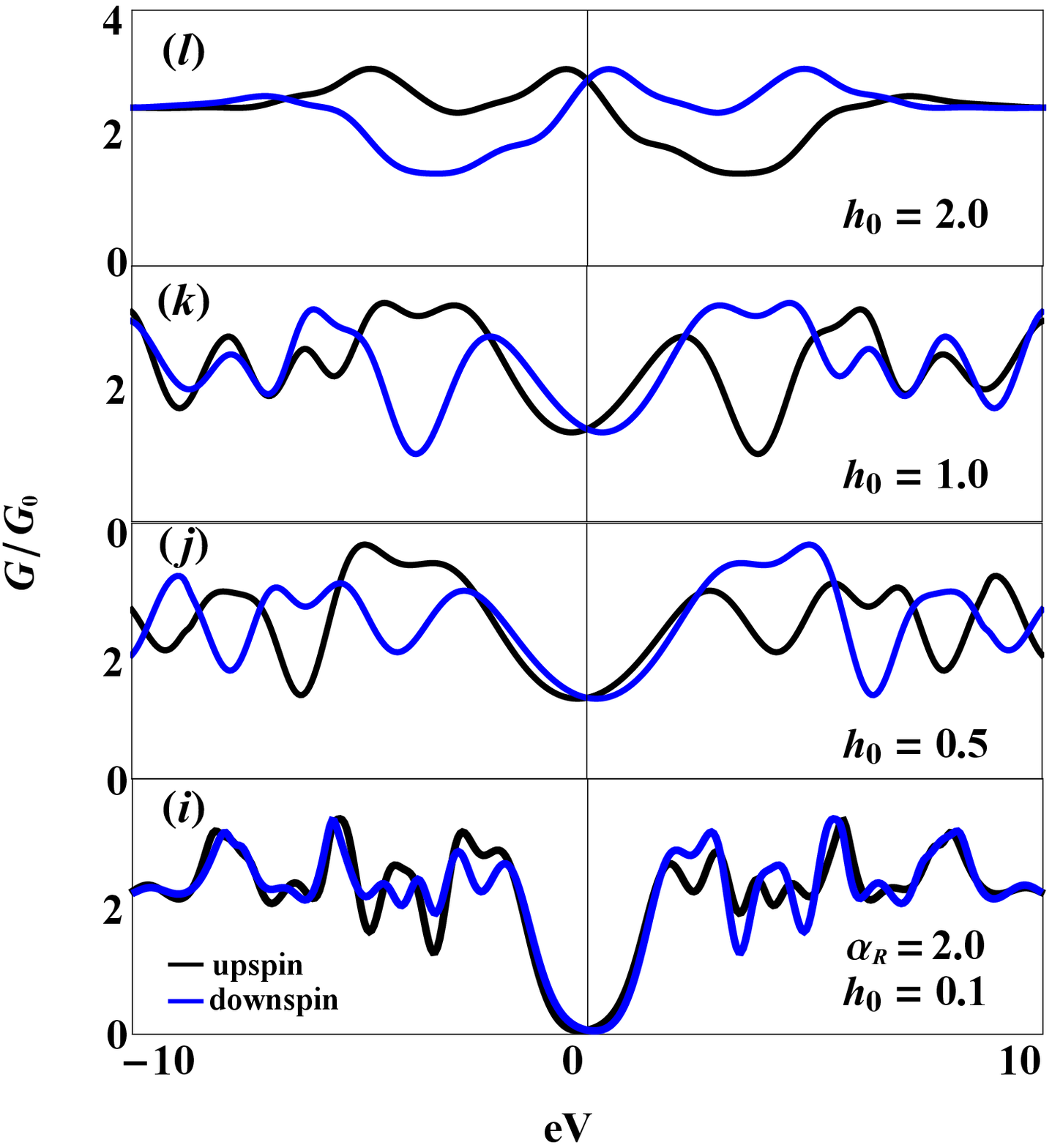}
\hspace{0.3cm}
}
\caption{Variation of conductance $G/G_0$ with $\text{eV}$ for different $h_0$ and $\alpha_{\text{R}}$ considering $\zeta = 0.3\pi$ and $\phi = 0.3\pi$. The plots (a)-(d) in the left panel are $\alpha_{\text{R}} = 0.1$, the plots (e)-(h) in the middle panel are for $\alpha_{\text{R}} = 1.0$ while the plots (i)-(l) in the right panel are for $\alpha_{\text{R}} = 2.0$ respectively.}
\label{fig7}
\end{figure*}

The variation of conductance with the bias voltage $\text{eV}$ of the proposed geometry is studied in Fig. \ref{fig7} for different choices of $\alpha_{\text{R}}$ and $h_0$ considering $\zeta = 0.3\pi$ and $\phi = 0.3\pi$. The plot in the left panel of Fig. \ref{fig7} is for $\alpha_{\text{R}} = 0.1$ while the plots in the middle and right panel are for $\alpha_{\text{R}} = 1.0$ and $2.0$ respectively. The conductance shows an oscillatory rise in low biasing region, for $\alpha_{\text{R}} = 0.1$ and $h_0 = 0.1$ but it saturates soon for moderate and higher $\text{eV}$ values. Notably, the conductance for up and down spin configurations is similar in this regime, as seen from Fig. \ref{fig7}(a). However, for $h_0 = 0.5$, the conductance is significantly different for both up and down spin configuration, as seen from Fig. \ref{fig7}(b). In this condition, the conductance minima get shifted to the respective sides for up and down spin electrons. It shows a linear rise in low biasing whilst an oscillatory response is observed with the rise in biasing energy and then saturates for strong biasing conditions. It follows the results of previously studied graphene systems \cite{masir,azarova,wang22}. It is noted that the conductance for both the up spin and down spin electrons are notably different in moderate and high $h_0$ systems. As soon as $h_0$ approaches $\text{E}_\text{F}$, the electrons acquire minimum energy from the Zeeman field and hence result in non-vanishing conductance even at zero bias condition as seen from Fig. \ref{fig7}(c). With the further rise in $h_0$ to $2$, the oscillations decrease and conductance of the order of $\sim 1.5G_0$ is found even at zero bias condition as seen from Fig. \ref{fig7}(d).
\begin{figure*}[hbt]
\centerline
\centerline{
\includegraphics[scale = 0.53]{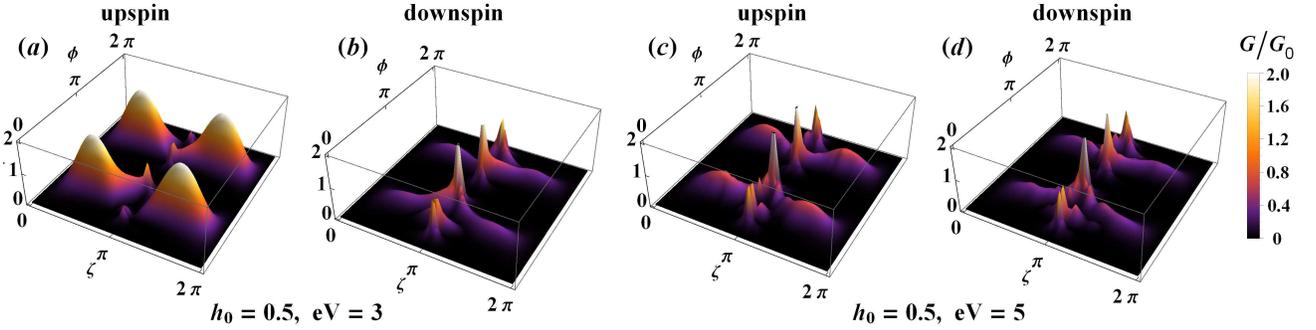}
\hspace{-0.1cm}
}
\caption{Variation of conductance $G/G_0$ with magnetization orientation $\zeta$ and $\phi$ for different biasing energy, considering $h_0 = 0.5$ and $\alpha_{\text{R}} = 1.0$. The plots (a) and (b) in the left panel are for $\text{eV} = 3$  while the plots (c) and (d) in the right panel are for $\text{eV} = 5$.}
\label{fig8}
\end{figure*} 
\begin{figure}[hbt]
\centerline
\centerline{
\includegraphics[scale = 0.5]{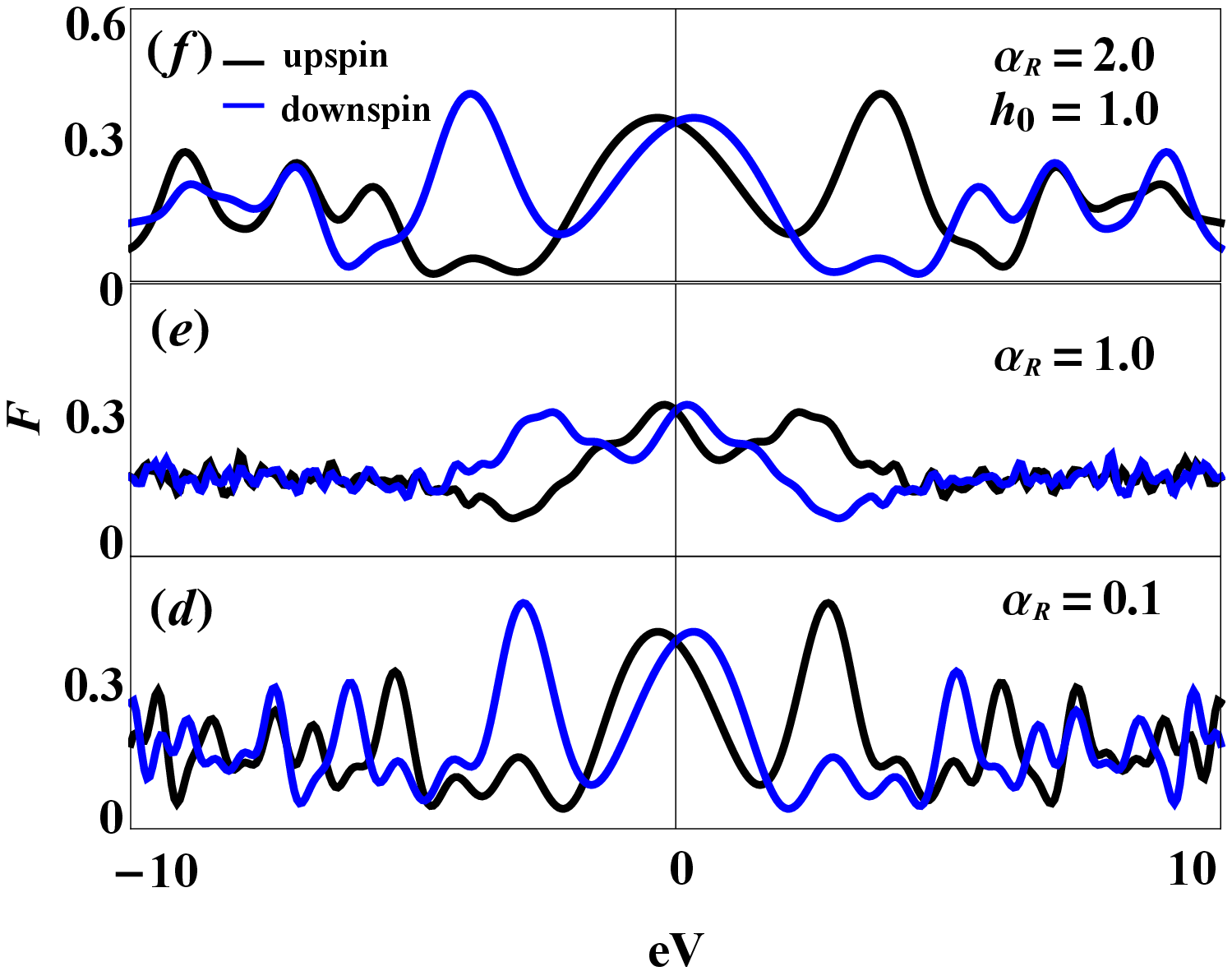}
\hspace{0.3cm}
\includegraphics[scale = 0.5]{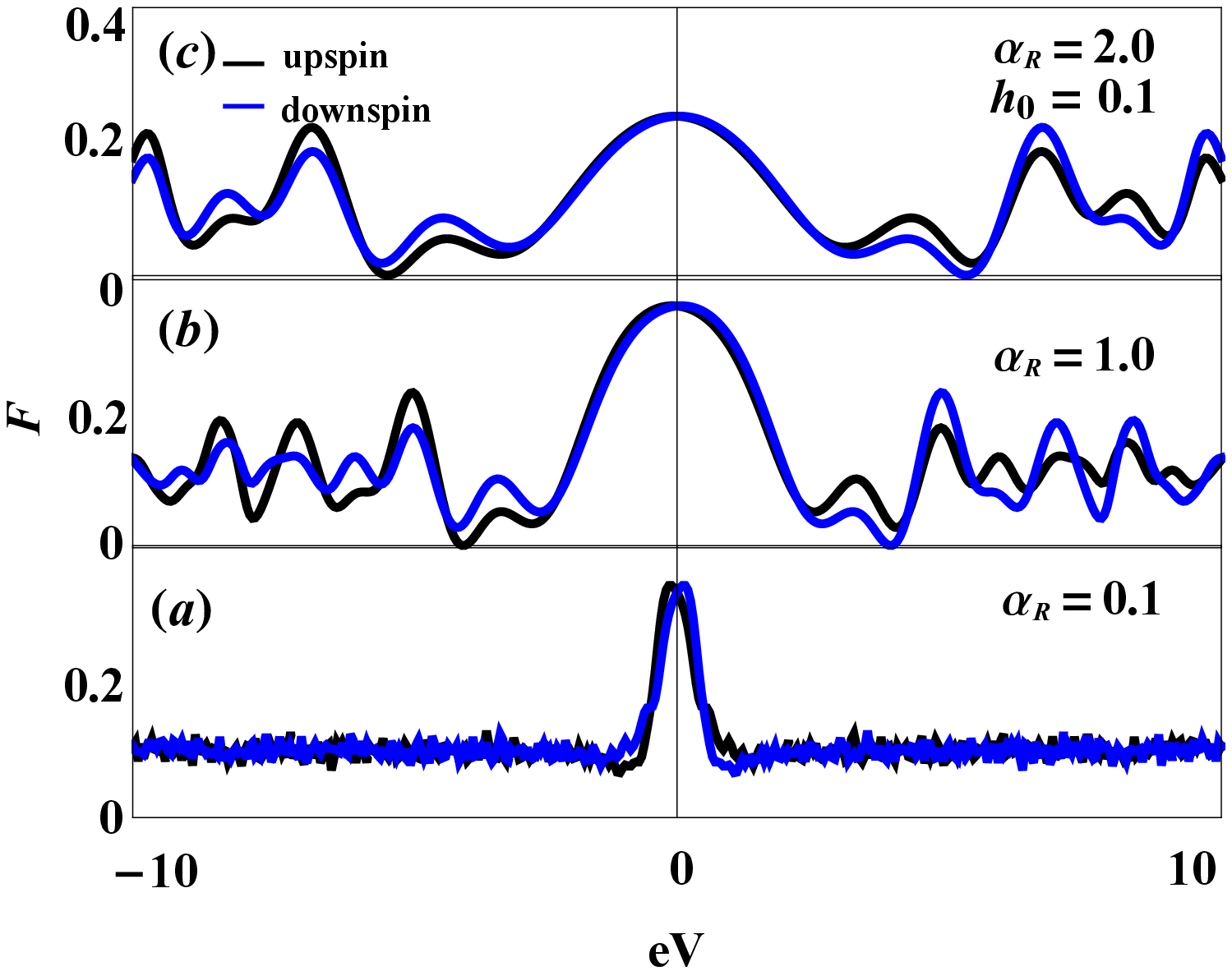}
}
\caption{Variation of Fano factor (F) with $\text{eV}$ for different choices of $h_0$ and $\alpha_{\text{R}}$ considering $\zeta = 0.3\pi$ and $\phi = 0.3\pi$. The plots are for (a) $h_0 = 1.0$, $\alpha_{\text{R}} = 0.1$ (b) $h_0 = 1.0$, $\alpha_{\text{R}} = 1.0$ (c) $h_0 = 1.0$, $\alpha_{\text{R}} = 2.0$.}
\label{fig9}
\end{figure}

With the increase in $\alpha_\text{R}$, we observe that the conductance gradually rises and then displays oscillatory behaviour for high biasing energy $\text{eV}$ with $h_0 = 0.1$  as seen from Figs. \ref{fig7}(e) and \ref{fig7}(i) respectively. This result follows the conductance found for graphene hybrids using magnetic arrangements \cite{masir}. But no significant change in up and down spin configuration is noticed in this condition. The conductance for up and down spin arrangements is different for $h_0 = 0.5$ as seen from Fig. \ref{fig7}(f). However, a notable difference is noticed for $\alpha_\text{R} = 2$. In this condition, the conductance is significantly increased even at low bias conditions, as seen from Fig. \ref{fig7}(j). It may be because the electron acquires sufficient energy for tunnelling from both Rashba and Zeeman fields. Similar characteristics are also seen in Figs. \ref{fig7}(g) and \ref{fig7}(k) with the further rise in $h_0$. But for $\alpha_\text{R} = 1$, the conductance grows initially and saturates with the rise in biasing energy as seen from Fig. \ref{eq7}(g). The conductance of the system display drastic change for $h_0 = 2$ in Figs. \ref{fig7}(h) and \ref{fig7}(l). In this condition, an oscillatory response of the system is observed as also observed for system with moderate $h_0$ and $\alpha_\text{R}$ values. It is to be noted that the conductance for moderate RSOC systems is $\sim 2G_0$ while for strong RSOC systems is $ \sim 3G_0$. A few notable observations are: (i) The conductance for up and down spin incoming electrons differ for moderate and high RSOC and magnetization strength. (ii) By tuning the magnetization and choosing a material with suitable RSOC, the conductance of RZ$|$BLG$|$RZ can enhance. (iii) There exists non-zero conductance even for low biasing conditions for both up and down spin incoming electrons, which is due to the presence of the Rashba - Zeeman effect.

Fig. \ref{fig8} shows the 3D plots of conductance $G/G_0$ with respect to $\zeta$ and $\phi$ for moderate magnetization ($h_0 = 0.5$) and with different biasing energy $\text{eV}$. We consider $\text{eV} = 3$ in plots \ref{fig8}(a) and \ref{fig8}(b) while $\text{eV} = 5$ in plots \ref{fig8}(c) and \ref{fig8}(d). We observe four broad maxima in conductance around the orientations $\zeta = \pi/2$ and $3\pi/2$ for up spin configurations while four sharp maxima are observed around the orientation $(\pi, 0.25\pi)$, $(\pi, 0.75\pi)$, $(\pi, 1.25\pi)$ and $(\pi, 1.75\pi)$ for down spin configuration at $\text{eV} = 3$ as observed from the plots \ref{fig8}(a) and \ref{fig8}(b). It is noted that the orientations $(\pi, 0.75\pi)$ and $(\pi, 1.25\pi)$ favour maximum conductance over other orientations for incoming down spin electrons. So, it can be concluded that at low bias voltage with moderate magnetization, the conductance of our system for up spin electrons will be due to the incoming electrons corresponding to the orientations of the broad maxima. But the orientations corresponding to the sharp maxima favours the incoming down spin electrons. With the increase in $\text{eV}$ to $5$, we see that the conductivity corresponding to the orientations $(\pi, 0.25\pi)$, $(\pi, 0.75\pi)$, $(\pi, 1.25\pi)$ and $(\pi, 1.75\pi)$ dominates over the orientation $\zeta = -\pi/2$ and $\pi/2$ for both up and down spin incoming electrons as seen from the plots \ref{fig8}(c) and \ref{fig8}(d) and also indicated from Figs. \ref{fig3} and \ref{fig4}. However, the intensity of the broad maxima for the up spin configuration is found to be more than that of the down spin configuration. So, it can be concluded that with the increase in biasing energy, the conductance of the up spin incoming electrons significantly changes more than that for down spin electrons. 

It is of our interest to investigate the shot noise for the RZ$|$BLG$|$RZ system. So in Fig. \ref{fig9}, we study the Fano factor $F$ with biasing voltage for different RSOC and $h_0$ considering $\zeta = 0.3\pi$ and $\phi = 0.3\pi$. We consider $h_0 = 0.1$ in plots \ref{fig9}(a), \ref{fig9}(b) and \ref{fig9}(c) while $h_0 = 1.0$ in plots \ref{fig9}(d), \ref{fig9}(e) and \ref{fig9}(f). The Fano factor shows a central maxima near $\text{eV} \sim 0$ for both up and down spin electron orientations and then display sharp decrease and saturates with increase in positive or negative bias voltage. It is to be noted that the maximum Fano factor for the system is found to be $\sim 0.3$ for this system. The peak near zero bias region broadens as $\alpha_\text{R}$ is increased to $1$ as seen from Fig. \ref{eq9}(b). However, the Fano factor shows an oscillatory behaviour with the increase in biasing energy. With the further rise in $\alpha_\text{R}$ to $2$, the central peak further broaden but the intensity of maximum Fano factor $F$ reduces and found to be $\sim 0.2$ in this condition. Moreover, the oscillations at higher bias voltage also become less frequent  as seen from Fig. \ref{eq9}(c)

Although the Fano factor for up and down configuration for low $h_0$ is not quite different, but as $h_0\approx\text{E}_\text{F}$ it is found to be significantly different for up and down spin configuration as seen from Fig. \ref{fig9}(d)-\ref{fig9}(f). However, in this case $F$ oscillates too frequently with the increase in bias voltage. The maxima of $F$ is found to be around $\text{eV} \sim \pm2.5$ for up and down spin configuration respectively. It should be noted that the maximum value of $F$ is found to be $\sim 0.37$ as seen from Fig. \ref{fig9}(d). The $F$ are notably reduced for $\alpha_\text{R} = 1$ for both up and down configuration. In this case, the oscillations are not prominently observed. However, the oscillations regains themselves with further rise in $\alpha_\text{R}$ to $2$ as observed from Fig. \ref{fig9}(f). Moreover, the maximum Fano factor in this condition is found to be $\sim 0.45$ around $\text{eV} = 0.4$.

\section{Conclusions}
In summary, we have studied spin-dependent ballistic transport and anomalous quantum tunnelling in RZ$|$BLG$|$RZ hybrid under external electric biasing. The primary purpose behind choosing such a hybrid geometry is the rapid growth of graphene-based hybrid devices in the recent era. We consider a double delta-like barrier in RZ$|$BLG$|$RZ geometry and an experimentally viable set of parameters for our study. We have investigated the role of various significant parameters like RSOC, strength and orientation of magnetization, applied bias voltage and the armchair dimension of the BLG system on transmission and conductance. Although the transmission for incoming up spin and down spin electrons are similar under low magnetization, they display notably different behaviour under moderate and high magnetization strength. The transmission is also asymmetric in the angular range of incidence and displays a set of resonances for moderate and high magnetization systems. One of the most astonishing results of our work is that the transmission of incoming up spin and down spin electrons are magnetization orientation dependent. Also, maximum tunnelling and thus conductance can be achieved by tuning some physically controllable parameters like biasing energy, magnetization strength and choosing a material with suitable RSOC. This unique property can be readily used in fabricating various devices such as spin filters which are in huge demand in the modern world. The Fano factor of our system is found to be $0.4$ for strong magnetization while reduces to $0.3$ under low magnetization conditions. Moreover, it can be reduced further with an appropriate choice of RSOC and bias energy. Furthermore, the transmission and also conductance has a strong dependence on RSOC. We observe that there exists a vital interplay of Rashba - Zeeman effect in tunnelling. The maximum transmission and conductance can be easily achieved by controlling magnetization strength for a low Rashba system at moderate biasing. This feature can be utilized to fabricate BLG-based hybrid structures considering an appropriate RZ material. Thus it provides a tremendous amount of control over the system and has a significant advantage in making stable devices.


\begin{thebibliography}{99}
\bibitem{novoselov11}
K. S. Novoselov, et. al., Science {\bf 306}, 666 (2004).
\bibitem{neto}
A. H. C. Neto, et. al., Rev. Mod. Phys. {\bf 81}, 109 (2009).
\bibitem{zhu}
Y. Zhu, et. al., Adv. Mater. {\bf 22}, 3906 (2010).
\bibitem{singh}
V. Singh, et. al., Prog. Mater. Sci. {\bf 56}, 1178 (2011).
\bibitem{jiang}
J. W. Jiang, J. S. Wang, and B. Li, B. Phys. Rev. B {\bf 80}, 113405 (2009).
\bibitem{lee}
C. Lee, X. Wei, J.W. Kysar, J. Hone, Science {\bf 321}, 5887 (2008).
\bibitem{mccann}
E. McCann and M. Koshino, Rep. Prog. Phys. {\bf 76}, 056503 (2013).
\bibitem{ghosh}
S. Ghosh, et. al., Appl. Phys. Lett. {\bf 92}, 151911, (2008).
\bibitem{balandin}
A. A. Balandin, et. al., Nano Lett. {\bf 8}, 902, (2008).
\bibitem{xu}
J. Xu, S. Dai, H. Li, and J. Yang,  New Carbon Mater. {\bf 33}, 213–220 (2018).
\bibitem{novoselov1}
K. S. Novoselov, et. al., Nature {\bf 438}, 197 (2005).
\bibitem{geim}
A. K. Geim and K. S. Novoselov, Nat. Mater. {\bf 6}, 183 (2007).
\bibitem{kymakis}
E. Kymakis, et. al., Thin Solid Films {\bf 520}, 1238–1241 (2011).
\bibitem{lin}
Y. M. Lin, et. al.,  Science {\bf 332}, 344017 (2011).
\bibitem{mattevi}
C. Mattevi, et. al., Nanotechnology {\bf 23}, 344017 (2012)
\bibitem{jangid}
P. Jangid, D. Pathan, and A. Kottantharayil, Carbon N.Y. {\bf 132}, 65–70 (2018).
\bibitem{liu1}
C. Liu, et. al., Nano Lett. {\bf 10}, 4863 (2010).
\bibitem{tan}
Y. B. Tan and J. M. Lee, J. Mater. Chem. A {\bf 1}, 14814 (2013).
\bibitem{purkait}
T. Purkait, et. al., Sci. Rep. {\bf 8}, 640 (2018).
\bibitem{brownson}
D. A. Brownson, D. K. Kampouris and C. E. Banks, Chem. Soc. Rev. {\bf 41} 6944, (2012).
\bibitem{wu1}
J. Wu, et. al., Sens. Actuators B {\bf 255}, 1805–1813 (2018).
\bibitem{pena}
J. Pe\~{n}a Bahamonde, H. N. Nguyen, S. K. Fanourakis and D. F. Rodrigues, J. Nanobiotechnol {\bf 16}, 75 (2018).
\bibitem{xu1}
W. Xu, et. al., Carbon N. Y. {\bf 141}, 247–252 (2019).
\bibitem{dean}
C. R. Dean, et. al., Nature Nanotechnol. {\bf 5}, 722 (2010).
\bibitem{zhang1}
J. Zhang and J Zhao, J. Appl. Phys. {\bf 113}, 043514 (2013).
\bibitem{gosling}
J. H. Gosling, et. al., Nat. Commun. Phys. {\bf 4}, 30 (2021).
\bibitem{tran}
T. H. Tran, et. al., Carbon, {\bf 176},  431-439 (2021).
\bibitem{novoselov}
K. S. Novoselov, et. al., Science {\bf 306}, 666 (2004).
\bibitem{novoselov2}
K. S. Novoselov, et. al., Nature Phys. {\bf 2}, 177 (2006).
\bibitem{ohta2}
T. Ohta, et. al., Science {\bf 313}, 951 (2006).
\bibitem{liang}
X. Liang, et. al., Phys. Rev. B {\bf 102}, 155146 (2020).
\bibitem{rozhkov}
A.V. Rozhkov,et. al., Phys. Rep. {\bf 648}, 1-104 (2016).
\bibitem{bagchi}
S. Bagchi, H. T. Johnson and H. B. Chew, Phys. Rev. B {\bf 101}, 054109 (2020).
\bibitem{song}
Z. Song, et. al., Phys. Rev. B {\bf 103}, 205412 (2021).
\bibitem{alavirad}
Y. Alavirad and J. Sau, Phys. Rev. B {\bf 102}, 235123 (2020).
\bibitem{santos}
H. Santos, A. Ayuela, L. Chico and E. Artacho, Phys. Rev. B {\bf 85}, 245430 (2012).
\bibitem{hu}
C. H. Hu, et. al., J. Phys. Chem. C {\bf 117}, 3572–3579 (2013).
\bibitem{mccann2}
E. McCann and V. I. Fal'ko, Phys. Rev. Lett. {\bf 96}, 086805 (2006).
\bibitem{guinea}
F. Guinea, A. H. Castro Neto and N. M. R. Peres, Phys. Rev. B {\bf 73}, 245426 (2006).
\bibitem{ohta31}
T. Ohta, et. al., Science {\bf 313}, 951 (2006).
\bibitem{oostinga}
J. B. Oostinga, et. al., Nature Mater. {\bf 7}, 151 (2007).
\bibitem{castro}
E. V. Castro, et. al., Phys. Rev. Lett. {\bf 99} 216802 (2007).
\bibitem{gorbachev}
R. V. Gorbachev, et. al., Phys. Rev. Lett. {\bf 98}, 176805 (2007).
\bibitem{morozov}
S. V. Morozov, et. al., Phys. Rev. Lett. {\bf 100}, 016602 (2008).
\bibitem{feldman}
B. E. Feldman, J. Martin and A. Yacoby, Nature Phys. {\bf 5}, 889 (2009).
\bibitem{xiao21}
S. Xiao, et. al., Phys. Rev. B {\bf 82} 041406 (2010).
\bibitem{koshino21}
M. Koshino and T. Ando Phys. Rev. B {\bf 73}, 245403 (2006).
\bibitem{cserti}
J. Cserti, Phys. Rev. B {\bf 75}, 033405 (2007).
\bibitem{snyman}
I. Snyman and C. W. J. Beenakker, Phys. Rev. B {\bf 75}, 045322 (2007).
\bibitem{cserti2}
J. Cserti and A. Csordas and G. David, Phys. Rev. Lett. {\bf 99}, 066802 (2007).
\bibitem{katsnelson}
M. I. Katsnelson, Eur. Phys. J. B {\bf 52}, 151 (2006).
\bibitem{katsnelson2}
M. I. Katsnelson, Phys. Rev. B {\bf 76}, 073411 (2007).
\bibitem{katsnelson3}
M. I. Katsnelson, K. S. Novoselov and A. K. Geim,  Nature Phys. {\bf 2} 620 (2007).
\bibitem{azarova}
E. S. Azarova and G. M. Maksimova, Physica E {\bf 61}, 118–124 (2014).
\bibitem{liu11}
J.-F. Liu, K.S. Chan and J. Wang, Nanotechnology {\bf 23} 095201 (2012).
\bibitem{zhang111}
Q. Zhang, Z.J. Lin and K.S. Chan, Appl. Phys. Lett. {\bf 102}, 142407 (2013).
\bibitem{chico}
L. Chico, A. Latge and L. Brey, Phys. Chem. Chem. Phys. {\bf 17}, 16469 (2015).
\bibitem{ganguly1}
S. Ganguly, S. Basu and S.K. Maiti, Europhys. Lett. {\bf 124} 57003 (2018).
\bibitem{ganguly2}
S. Ganguly, S. Basu, S.K. Maiti, Superlattices Microstruct. {\bf 120}, 650 (2018).
\bibitem{fouladi1}
A. A. Fouladi, Physica E, {\bf 102}, 117-122 (2018).
\bibitem{zhang121}
Q. Zhang, J. Jiang and K.S.Chan, Physics Letters A {\bf 383} 2957–2962 (2019).
\bibitem{liu111}
J.-H. Liu, et. al., RSC Adv., {\bf 12}, 3386 (2022).
\bibitem{rashba}
L. P. Gor’kov, E. I. Rashba, Phys. Rev. Lett. {\bf 87}, 037004 (2001).
\bibitem{lashell}
S. LaShell, B. A. McDougall and E. Jensen, Phys. Rev. Lett. {\bf 77}, 3419 (1996).
\bibitem{acharjee1}
S. Acharjee and U. D. Goswami, Supercond. Sci. Technol. {\bf 32}, 085004 (2019).
\bibitem{acharjee2}
S. Acharjee and U. D. Goswami, J. Magn. Magn. Mater. {\bf 495}, 165844 (2020).
\bibitem{acharjee3}
S. Acharjee and U. D. Goswami, Physica E {\bf 135}, 114967 (2022).
\bibitem{cavigilia}
A. D. Caviglia,  et al., Phys. Rev. Lett. {\bf 104}, 126803 (2010).
\bibitem{ishizaka}
K. Ishizaka, et. al., Nat. Mater. {\bf 10}, 521–526 (2011).
\bibitem{shinova}
J. Sinova, et. al., Phys. Rev. Lett. {\bf 92}, 126603 (2004).
\bibitem{tao1}
L. Tao and E. Y. Tsymbal, npj Comput Mater {\bf 6}, 172 (2020). 
\bibitem{xiao}
X. Xiao, et. al., J. App. Phys. {\bf 115}, 223709 (2014).
\bibitem{zhai}
L. X. Zhai, Y. Wang, and Z. An, AIP Advances {\bf 8}, 055120 (2018).
\bibitem{krempasky}
J. Krempasky, et. al., Nat. Commun. {\bf 7}, 13071 (2016).
\bibitem{kammhuber}
J.	Kammhuber, et. al., Nat. Commun. {\bf 8}, 478 (2017).
\bibitem{rozhkov1}
A. V. Rozhkov, A. O. Sboychakov, A. L. Rakhmanov and F. Nori, Phys. Rep. {\bf 648}, 1-104 (2016).
\bibitem{abdullah}
H. M. Abdullah, M. A. Ezzi and H. Bahlouli, J. App. Phys. {\bf 124}, 204303 (2018).
\bibitem{masir}
M. R. Masir, P. Vasilopoulos and F. M. Peeters, Phys. Rev. B {\bf 79}, 035409 (2009).
\bibitem{mccann3}
E. McCann, Phys. Rev. B {\bf 74}, 161403(R) (2006).
\bibitem{min}
H. Min, B. R. Sahu, S. K. Banerjee and A. H. MacDonald, Phys. Rev. B {\bf 75}, 155115 (2007).
\bibitem{nilsson}
J. Nilsson, A. H. Neto Castro, F. Guinea and N. M. R. Peres, Phys. Rev. B {bf 78}, 045405 (2008).
\bibitem{zhang11}
L. M. Zhang, et. al., Phys. Rev. B {\bf 78}, 235408 (2008).
\bibitem{li}
Z. Q. Li, et. al., Phys. Rev. Lett. {\bf 102}, 037403 (2009).
\bibitem{dresselhaus}
M. S. Dresselhaus and G. Dresselhaus, {\bf 51}, 1 (2002).
\bibitem{mucha}
M. Mucha-Kruczy\'{n}ski, E. Mccann and V. I. Fal'ko, Semicond. Sci. Technol. {\bf 25}, 033001 (2010).
\bibitem{linder}
J. Linder and A. Sudb\o{}, Phys. Rev. B {\bf 75}, 134509 (2007).\
\bibitem{lejarreta}
J. D. Lejarreta, C. H. Fuentevilla, E. Diez and J. M. Cerver\'{o}, J. Phys. A: Math. Theor. {\bf 46}, 155304 (2013).
\bibitem{buttiker}
Ya. M. Blanter and M. B\"{u}ttiker, Phys. Rep. {\bf 336}, 1 (2000).
\bibitem{wang22}
Y. Wang, J. Appl. Phys. {\bf 116}, 164317 (2014).
\bibitem{zhang22}
Q. Zhang, J. Jiang and K. S. Chan, Phys Lett. A. {\bf 383}, 2957-2962 (2019).
\end{thebibliography}
\end{document}